\documentclass[aps,prd,onecolumn,showpacs,superscriptaddress,nofootinbib]{revtex4-2}

\usepackage{amsmath,amssymb,mathtools}
\usepackage{hyperref}
\usepackage{xcolor}
\usepackage{tikz}
\newcommand{\Dbar}{\bar D}
\usetikzlibrary{calc,decorations.pathmorphing,decorations.markings}

\tikzset{
  bgrav/.style={
    thick,decorate,
    decoration={coil,aspect=0.65,segment length=4.8pt,amplitude=3pt}
  },
  qgrav/.style={
    thick,decorate,
    decoration={snake,amplitude=1.4pt,segment length=7pt}
  },
  qlm/.style={thick,dashed},
  bscalar/.style={thick,double,double distance=1.4pt},
  qscalar/.style={thick,line cap=round},
  qslm/.style={thick,line cap=round,
    dash pattern=on 5pt off 2.5pt on 0.8pt off 2.5pt},
  ghost/.style={thick,densely dotted}
}

\newcommand{\propfig}[2]{%
\begin{tikzpicture}[baseline=-0.5ex,line cap=round,line join=round]
  \draw[#1] (0,0)--(2.2,0);
\end{tikzpicture}
\;=\; #2
}

\newcommand{\mixedWDc}[2]{%
  \draw[qgrav] (#1) -- ($ (#1)!0.5!(#2) $);
  \draw[qlm]   ($ (#1)!0.5!(#2) $) -- (#2);
}
\newcommand{\mixedDWc}[2]{%
  \draw[qlm]   (#1) -- ($ (#1)!0.5!(#2) $);
  \draw[qgrav] ($ (#1)!0.5!(#2) $) -- (#2);
}

\newcommand{\mixedWD}[4]{%
  \draw[qgrav] (#1,#2) -- ($(#1,#2)!0.5!(#3,#4)$);
  \draw[qlm]   ($(#1,#2)!0.5!(#3,#4)$) -- (#3,#4);
}

\newcommand{\mixedSscalar}[4]{%
  \draw[qscalar] (#1,#2) -- ($(#1,#2)!0.5!(#3,#4)$);
  \draw[qslm]    ($(#1,#2)!0.5!(#3,#4)$) -- (#3,#4);
}
\newcommand{\mixedSscalarc}[2]{%
  \draw[qscalar] (#1) -- ($ (#1)!0.5!(#2) $);
  \draw[qslm]    ($ (#1)!0.5!(#2) $) -- (#2);
}
\newcommand{\mixedSscalarRevc}[2]{%
  \draw[qslm]    (#1) -- ($ (#1)!0.5!(#2) $);
  \draw[qscalar] ($ (#1)!0.5!(#2) $) -- (#2);
}

\newcommand{\SelfEnergyCoords}{%
  \coordinate (Lext) at (-3.0,0);
  \coordinate (L)    at (-1.4,0);
  \coordinate (R)    at ( 1.4,0);
  \coordinate (Rext) at ( 3.0,0);
  \coordinate (Cup)  at ( 0.0, 1.1);
  \coordinate (Cdw)  at ( 0.0,-1.1);
}

\newcommand{\UpperLeft}[1]{%
  \draw[#1] (L) arc[start angle=180,end angle=90,
                    x radius=1.4,y radius=1.1];
}
\newcommand{\UpperRight}[1]{%
  \draw[#1] (Cup) arc[start angle=90,end angle=0,
                      x radius=1.4,y radius=1.1];
}
\newcommand{\LowerLeft}[1]{%
  \draw[#1] (L) arc[start angle=180,end angle=270,
                    x radius=1.4,y radius=1.1];
}
\newcommand{\LowerRight}[1]{%
  \draw[#1] (Cdw) arc[start angle=270,end angle=360,
                      x radius=1.4,y radius=1.1];
}
\newcommand{\Vertices}{%
  \fill (L) circle (1.3pt);
  \fill (R) circle (1.3pt);
}
\newcommand{\vertexdot}[1]{\fill (#1) circle (1.25pt);}

\newcommand{\threevertex}[6]{%
\begin{tikzpicture}[baseline=-0.5ex,line cap=round,line join=round]
  \coordinate (C) at (0,0);
  \draw[#1] (C)--(-1.2,0.9);
  \draw[#2] (C)--(-1.2,-0.9);
  \draw[#3] (C)--(1.35,0);
  \fill (C) circle (1.3pt);
  \node[left]  at (-1.25, 0.9) {$#4$};
  \node[left]  at (-1.25,-0.9) {$#5$};
  \node[right] at ( 1.4,  0  ) {$#6$};
\end{tikzpicture}
}
\newcommand{\fourvertex}[8]{%
\begin{tikzpicture}[baseline=-0.5ex,line cap=round,line join=round]
  \coordinate (C) at (0,0);
  \draw[#1] (C)--(-1.2, 1.0);
  \draw[#2] (C)--(-1.2,-1.0);
  \draw[#3] (C)--( 1.2, 1.0);
  \draw[#4] (C)--( 1.2,-1.0);
  \fill (C) circle (1.3pt);
  \node[left]  at (-1.25, 1.0) {$#5$};
  \node[left]  at (-1.25,-1.0) {$#6$};
  \node[right] at ( 1.25, 1.0) {$#7$};
  \node[right] at ( 1.25,-1.0) {$#8$};
\end{tikzpicture}
}

\numberwithin{equation}{section}
\renewcommand{\theequation}{\arabic{section}.\arabic{equation}}

\hypersetup{
  colorlinks=true,
  linkcolor=blue,
  citecolor=blue,
  urlcolor=blue
}

\begin{document}

\title{Renormalization of Quantum Gravity with Lagrange Multiplier Fields}

\author{F.~T.~Brandt}
\email{fbrandt@usp.br}
\affiliation{Instituto de F\'{\i}sica, Universidade de S\~ao Paulo,
  S\~ao Paulo, SP 05508-090, Brazil}

\author{J.~Frenkel}
\email{jfrenkel@if.usp.br}
\affiliation{Instituto de F\'{\i}sica, Universidade de S\~ao Paulo,
  S\~ao Paulo, SP 05508-090, Brazil}

\author{S.~Martins-Filho}
\email{s.martins-filho@unesp.br}
\affiliation{Instituto de F\'{\i}sica Te\'orica, Universidade Estadual
  Paulista,  S\~ao Paulo, SP 01140-070, Brazil}

\author{D.~G.~C.~McKeon}
\email{dgmckeo2@uwo.ca}
\affiliation{Department of Applied Mathematics,
  The University of Western Ontario,
  London, Ontario N6A~5B7, Canada}
\affiliation{Department of Mathematics and Computer Science,
  Algoma University,
  Sault Ste.~Marie, Ontario P6A~2G4, Canada}

\date{\today}


\begin{abstract}
We study the renormalization of Einstein gravity coupled to
a scalar field in the Lagrange multiplier (LM) formulation,
where a pair of quantum LM fields, $(\lambda'_{\mu\nu},\sigma')$,
is coupled to the gauge-fixed gravitational and scalar equations of
motion. Using the fact that in this approach there are no higher loop corrections,
we show that renormalizing the two quantum LM fields at one-loop order absorbs,
off-shell, five out of the six independent divergent structures. The
remaining off-shell counterterm, proportional to
\(
(\partial_\mu\bar\phi\,\partial^\mu\bar\phi)^2,
\)
is nonzero on shell and cannot be absorbed by any LM field
redefinition, since such redefinitions generate only counterterms that
vanish on shell. In contrast, in pure gravity (no matter), the LM renormalization
absorbs all one-loop divergences and no additional counterterms are
required. The LM formulation therefore provides a sharp
off-shell separation between equation-of-motion divergences, absorbed
into LM field renormalizations, and the single irreducible structure.
\end{abstract}

\maketitle

\section{Introduction}
\label{sec:intro}

The quantization of gravity remains one of the central open problems
in theoretical physics. Einstein's theory of general relativity,
when treated as a quantum field theory, is perturbatively
non-renormalizable: the loop expansion generates an infinite
sequence of ultraviolet divergences requiring an infinite number
of independent counterterms; for pure gravity the first genuine
on-shell divergence appears at two loops~\cite{Goroff:1985sz,Goroff:1985th}.
At one loop, it was shown in Ref.~\cite{tHooft:1974toh} that pure
quantum gravity is on-shell renormalizable (its divergences can
be removed by a field redefinition), while coupling to a scalar
field breaks this property: a genuine on-shell divergence arises
proportional to $(\partial_\mu\bar\phi\,\partial^\mu\bar\phi)^2$,
which cannot be absorbed by any redefinition of the original
fields. This result was confirmed and extended to a general
two-parameter background gauge in Ref.~\cite{Frenkel:2026zwc},
where the gauge-independence of the on-shell effective action
predicted by the DeWitt--Kallosh theorem~\cite{DeWitt:1967yk,Kallosh:1974yh}
was also verified explicitly. The current point of view is that
general relativity should be regarded as a low-energy effective
field theory~\cite{Donoghue:1995}, valid below some fundamental
scale, and it remains of theoretical interest to identify
formulations that simplify or reorganize the structure of its
ultraviolet divergences~\cite{Brandt:2023uaw}.

One of the aims of the present work is to show that the Lagrange
multiplier (LM) formulation provides a sharp separation between the
removable off-shell divergences and the genuine irremovable
ultraviolet singularities at one-loop order. As shown in
Refs.~\cite{McKeon:1992,Chishtie:2013,Brandt:2019ymg,Brandt:2020cjp,Brandt:2021aop,McKeon:2021rle,McKeon:2026liv,Brandt:2023lyq,McKeon:2024psy},
the LM formulation of quantum gravity leads to a unitary theory, in
which there are no higher-loop contributions beyond the one-loop order.

In this formulation, a purely quantum tensor field
$\lambda'_{\mu\nu}$ (quantum part of the split
$\lambda_{\mu\nu} = \bar\lambda_{\mu\nu} + \lambda'_{\mu\nu}$,
with $\bar\lambda_{\mu\nu}=0$) is coupled
linearly to the gravitational equation of motion. The ordinary
graviton propagator $\langle hh\rangle_q$ vanishes identically,
and internal gravitational lines are replaced by the mixed
propagators $\langle h\lambda'\rangle$, $\langle\lambda' h\rangle$,
and $\langle\lambda'\lambda'\rangle$. This reorganizes the
perturbative expansion and removes entire classes of diagrams
present in the ordinary formulation; one-loop
exactness holds and all
higher-loop radiative corrections vanish
identically~\cite{Brandt:2023lyq,McKeon:2024psy}. This holds for
$\lambda'_{\mu\nu}$ coupled to the pure gravitational sector.

Once matter is included, however, the tensor LM field
$\lambda'_{\mu\nu}$ alone is no longer sufficient to eliminate all
higher-loop contributions. As shown in Sec.~\ref{sec:puregravity}
and illustrated explicitly at two-loop order in
Appendix~\ref{app:twoloop}, an LM partner for the scalar field is also
required. We therefore introduce a second purely quantum LM field
$\sigma'$
(with $\sigma = \bar\sigma + \sigma'$, $\bar\sigma=0$) for the scalar
sector. In the gauge-fixed background-field formulation of
Ref.~\cite{Frenkel:2026zwc}, the gauge-fixing function contains
the quantum scalar field $\varphi$ through a term proportional to
$\zeta\kappa\varphi\,\partial^\mu\bar\phi$. The natural LM
construction therefore couples both $\lambda'_{\mu\nu}$ and a
scalar LM field $\sigma'$ to the \emph{gauge-fixed} equations of
motion:
\begin{equation*}
S_{\rm LM}^{\rm tot}
=
\int d^4x\left[
\lambda'_{\mu\nu}\frac{\delta(S_{\rm inv}+S_{\rm gf})}{\delta h_{\mu\nu}}
+
\sigma'\frac{\delta(S_{\rm inv}+S_{\rm gf})}{\delta\varphi}
\right].
\end{equation*}
This guarantees that the quadratic LM matrix takes the canonical
block form with the same gauge-fixed operator in both the ordinary
and mixed blocks, so that $\langle\varphi\varphi\rangle_q=0$ as
well. The determinant doubling produced by this block structure is
cancelled by Lee--Yang ghost fields acting on the full coupled
gauge-fixed Hessian, establishing the equivalence of the LM and
ordinary one-loop counterterms.

We then ask whether renormalizing both LM fields can absorb all
one-loop divergences of the theory, rendering it finite at one
loop without additional counterterms. Our results are as follows.

The tensor LM renormalization of $\lambda'_{\mu\nu}$,
Eq.~\eqref{eq:lambda-renorm}, absorbs the four independent
gravitational and mixed counterterm structures
$\bar R_{\mu\nu}\bar R^{\mu\nu}$, $\bar R^2$,
$\bar R\,\partial_\mu\bar\phi\,\partial^\mu\bar\phi$, and
$\bar R_{\mu\nu}\partial^\mu\bar\phi\,\partial^\nu\bar\phi$,
fixing four free parameters $c_1,\ldots,c_4$ uniquely. The
scalar LM renormalization of $\sigma'$,
Eq.~\eqref{eq:sigma-renorm}, absorbs the remaining
equation-of-motion structure $\kappa^2\mathcal{C}_6(\Dbar^2\bar\phi)^2$,
which vanishes on shell and is present only when $\zeta\neq0$.
This leaves a single unmatched structure,
\begin{equation*}
\frac{203}{320}\,\kappa^4
\bigl(\partial_\mu\bar\phi\,\partial^\mu\bar\phi\bigr)^2,
\end{equation*}
which is nonzero on shell and gauge-independent. Thus the LM
formulation identifies the same physical obstruction as the ordinary
treatment~\cite{tHooft:1974toh,Frenkel:2026zwc}, but only after all
off-shell equation-of-motion structures have been absorbed into LM
field renormalizations.

The structural reason this obstruction cannot be removed is clear:
both $\lambda'_{\mu\nu}$ and $\sigma'$ enter the LM action linearly,
multiplied by expressions that vanish on the respective mass
shells. Therefore any local renormalization of $\lambda'_{\mu\nu}$
or $\sigma'$ generates only equation-of-motion counterterms, which
vanish on shell and can never cancel a genuine on-shell divergence.
A striking contrast arises with pure gravity: there, all one-loop
divergences are curvature-squared terms that are equation-of-motion
counterterms (proportional to the vacuum Einstein equations), and
the LM renormalization absorbs them all with no residual. When a
scalar field is coupled, the counterterm acquires the nonzero on-shell
value $\kappa^4(203/320)(\partial\bar\phi)^4$, which no
equation-of-motion counterterm can cancel.
We use the shorthand
$(\partial\bar\phi)^4 \equiv (\partial_\mu\bar\phi\,\partial^\mu\bar\phi)^2$
throughout.

The paper is organized as follows. In Sec.~\ref{sec:formulation}
we present the gauge-fixed LM formulation of gravity coupled to
scalar matter, including the background-field decomposition, the
general two-parameter gauge fixing, the propagators, and the
vertex rules. In Sec.~\ref{sec:CT} we draw and identify the
one-loop LM diagrams, introduce the LM renormalization ansätze,
derive the divergent Lagrangians, and match against the full
one-loop counterterm to determine all renormalization coefficients.
In Sec.~\ref{sec:obstruction} we analyse the residual obstruction,
establish its gauge-independence, and give the structural argument.
In Sec.~\ref{sec:puregravity} we contrast pure gravity and the
scalar-coupled case. Section~\ref{sec:conclusion} contains our
conclusions. Appendix~\ref{app:diagrams} collects the LM Feynman
rules. Appendix~\ref{app:matching} gives the algebraic verification
of the matching and the gauge-independence proof. Appendix~\ref{app:phi3}
illustrates the structural argument of Sec.~\ref{sec:CT} in a simpler
$\phi^3$ toy model with a Lagrange multiplier field. Appendix~\ref{app:twoloop}
gives explicit two-loop diagrams, and a general topological argument,
showing why both LM fields are required once matter is present.

We use the metric signature $(+,-,-,-)$ and work in $D=4-2\epsilon$
spacetime dimensions with dimensional regularization. The coupling
constant is $\kappa^2=16\pi G_N$. Our Riemann and Ricci tensor sign
conventions follow those of Refs.~\cite{tHooft:1974toh,Frenkel:2026zwc}.

\section{The LM formulation of quantum gravity coupled to scalar matter}
\label{sec:formulation}

\subsection{The classical action}

We consider a scalar field $\phi$ minimally coupled to
gravity, described by the 't Hooft-Veltman  action \cite{tHooft:1974toh}
\begin{equation}
S[g_{\mu\nu},\phi]
=
S_g[g_{\mu\nu}] + S_\phi[g_{\mu\nu},\phi],
\label{eq:S-total}
\end{equation}
with the Einstein--Hilbert and scalar parts given by
\begin{equation}
S_g[g_{\mu\nu}]
=
\frac{1}{\kappa^2}
\int d^4x\,\sqrt{-g}\,R,
\label{eq:Sg}
\end{equation}
\begin{equation}
S_\phi[g_{\mu\nu},\phi]
=
\frac{1}{2}\int d^4x\,\sqrt{-g}\,
g^{\mu\nu}\partial_\mu\phi\,\partial_\nu\phi,
\label{eq:Sphi}
\end{equation}
where $\kappa^2 = 16\pi G_N$ and $R$ is the Ricci scalar.
All Feynman-rule signs below follow the $(+,-,-,-)$ scalar-action
convention of Eq.~\eqref{eq:Sphi} and the conventions of
Ref.~\cite{Frenkel:2026zwc}.
The energy-momentum tensor of the scalar field is
\begin{equation}
T_{\mu\nu}
=
\frac{1}{2}g_{\mu\nu}\,\partial_\alpha\phi\,\partial^\alpha\phi
-
\partial_\mu\phi\,\partial_\nu\phi.
\label{eq:Tmunu}
\end{equation}
The gravitational and scalar equations of motion are, respectively,
\begin{equation}
R_{\mu\nu} - \frac{1}{2}Rg_{\mu\nu}
- \frac{\kappa^2}{2}T_{\mu\nu}
= 0,
\label{eq:EOM-grav}
\end{equation}
\begin{equation}
\mathrm{D}_\mu\mathrm{D}^\mu\phi
= 0,
\label{eq:EOM-scalar}
\end{equation}
where $\mathrm{D}_\mu$ denotes the covariant derivative with respect
to $g_{\mu\nu}$. It will be convenient to write the scalar
equation-of-motion operator as
\begin{equation}
E_\phi \equiv \frac{1}{\sqrt{-g}}\frac{\delta S_\phi}{\delta\phi}
= -\,\mathrm{D}_\mu\mathrm{D}^\mu\phi ,
\label{eq:scalar-eom-operator}
\end{equation}
which vanishes on the scalar mass shell, $E_\phi = 0$.

\subsection{Background-field decomposition and the LM actions}

We use the background-field method~\cite{Abbott:1981ke}, splitting
the fields into background and quantum parts:
\begin{equation}
g_{\mu\nu} = \bar g_{\mu\nu} + \kappa h_{\mu\nu},
\qquad
\phi = \bar\phi + \varphi.
\label{eq:split}
\end{equation}
Here $\bar g_{\mu\nu}$ and $\bar\phi$ are the background fields,
while $h_{\mu\nu}$ and $\varphi$ are the quantum fields that
propagate inside loops.

Following the LM formalism of
Refs.~\cite{Brandt:2019ymg,Brandt:2021aop,Brandt:2023lyq,McKeon:2024psy}, we
introduce a Lagrange multiplier field $\lambda_{\mu\nu}$, split into
background and quantum parts,
$\lambda_{\mu\nu} = \bar\lambda_{\mu\nu} + \lambda'_{\mu\nu}$,
with no background part:
\begin{equation}
\bar\lambda_{\mu\nu} = 0.
\label{eq:lambda-purely-quantum}
\end{equation}
The LM field is coupled to the gauge-fixed equations of motion
as described below. The invariant part of the tensor LM coupling is
proportional to the gravitational equation of motion
\eqref{eq:EOM-grav}:\footnote{The LM Lagrangian couples
$\lambda'_{\mu\nu}$ to the tensor combination
$G_{\mu\nu}/\kappa - \kappa T_{\mu\nu}/2$, which vanishes on shell.
The LM formalism of
Refs.~\cite{Brandt:2019ymg,Brandt:2021aop,Brandt:2023lyq,McKeon:2024psy}
uses a related but not identical construction; the tree-level
assignment $\bar\lambda_{\mu\nu}=0$ is specific to the present work.}
\begin{equation}
\mathcal{L}_{\rm LM}^{\rm inv}
=
-\sqrt{-g}\,\lambda'^{\mu\nu}
\left[
\frac{1}{\kappa}
\!\left(
R_{\mu\nu}-\frac{1}{2}Rg_{\mu\nu}
\right)
-
\frac{\kappa}{2}T_{\mu\nu}
\right],
\label{eq:LLM-inv}
\end{equation}
where $T_{\mu\nu}$ is the energy-momentum tensor \eqref{eq:Tmunu}.
This sign follows from the convention in which the LM field is coupled to
$\delta S_{\rm inv}/\delta h_{\mu\nu}$, while the metric fluctuation is
defined by $g_{\mu\nu}=\bar g_{\mu\nu}+\kappa h_{\mu\nu}$; varying with
respect to the covariant metric gives an extra minus sign. This is the
normalization convention used throughout the paper.
The full gauge-fixed LM action is specified below,
Eq.~\eqref{eq:SLMtot}.

In addition to the tensor field $\lambda'_{\mu\nu}$, we now introduce
a second, \emph{scalar} Lagrange multiplier field $\sigma$, split as
$\sigma = \bar\sigma + \sigma'$, also purely quantum,
\begin{equation}
\bar\sigma = 0.
\label{eq:sigma-purely-quantum}
\end{equation}
In the gauge-fixed background-field formulation, the natural LM
coupling is to the \emph{gauge-fixed} quantum equations of motion.
In field-space notation, with $Q^A=(h_{\mu\nu},\varphi)$ and
$\Lambda_A=(\lambda'_{\mu\nu},\sigma')$, the LM action is
\begin{equation}
S_{\rm LM}^{\rm tot}
=
\int d^4x\,\Lambda_A\,
\frac{\delta(S_{\rm inv}+S_{\rm gf})}{\delta Q^A}
=
\int d^4x\left[
\lambda'_{\mu\nu}\,\frac{\delta(S_{\rm inv}+S_{\rm gf})}{\delta h_{\mu\nu}}
+
\sigma'\,\frac{\delta(S_{\rm inv}+S_{\rm gf})}{\delta\varphi}
\right],
\label{eq:SLMtot}
\end{equation}
where $S_{\rm inv} = S_g+S_\phi$ and $S_{\rm gf}$ is the
gauge-fixing action specified below. This guarantees that the
quadratic LM matrix takes the canonical block form
\begin{equation}
K =
\begin{pmatrix} O & O \\ O & 0 \end{pmatrix},
\qquad
K^{-1} =
\begin{pmatrix} 0 & O^{-1} \\ O^{-1} & -O^{-1} \end{pmatrix},
\label{eq:LM-block}
\end{equation}
with the \emph{same} gauge-fixed operator $O$ in both the ordinary
and the mixed blocks. Varying $S_{\rm LM}^{\rm tot}$ with respect to
the multipliers gives the gauge-fixed equations $\delta(S_{\rm
inv}+S_{\rm gf})/\delta Q^A = 0$, which on shell reduce to the
classical equations of motion; the classical content of the theory is
unchanged.

A key property of the LM construction is \emph{one-loop exactness}: since
$\Lambda_A = (\lambda'_{\mu\nu},\sigma')$ appear linearly in
$S_{\rm LM}^{\rm tot}$, integrating them out in the path integral
produces delta functionals $\delta[\delta(S_{\rm inv}+S_{\rm
gf})/\delta Q^A]$ that enforce the quantum equations of motion.
The path integral thereby localizes on field configurations satisfying
those equations, and all quantum loop contributions beyond one loop
vanish identically~\cite{Brandt:2023lyq,McKeon:2024psy}.

The tensor LM coupling gives the invariant part
\begin{equation}
\mathcal{L}_{\rm LM}
=
-\sqrt{-g}\,\lambda'^{\mu\nu}
\!\left[
\frac{1}{\kappa}
\!\left(
R_{\mu\nu}-\frac{1}{2}Rg_{\mu\nu}
\right)
-
\frac{\kappa}{2}T_{\mu\nu}
\right]
+
\lambda'_{\mu\nu}\,\frac{\delta S_{\rm gf}}{\delta h_{\mu\nu}},
\label{eq:LLM}
\end{equation}
where $T_{\mu\nu}$ is the energy-momentum tensor \eqref{eq:Tmunu}.
The gauge-fixing descendant $\lambda'_{\mu\nu}\,\delta S_{\rm gf}/\delta h_{\mu\nu}$
is what ensures that the $h\lambda'$ block of the quadratic matrix
equals the full gauge-fixed graviton operator $O_{\mu\nu,\rho\sigma}$,
including the de~Donder term. The corresponding action is
\begin{equation}
S_{\rm LM}
=
\int d^4x\,\mathcal{L}_{\rm LM}.
\label{eq:SLM}
\end{equation}

For the scalar LM field, the invariant coupling is
$\sigma' E_\phi = -\sigma'\,\mathrm{D}^2\phi$
(Eq.~\eqref{eq:scalar-eom-operator}). Since the gauge-fixing
function $\chi^\mu$ of Ref.~\cite{Frenkel:2026zwc} contains the
quantum scalar field $\varphi$ through the term
$-\zeta\kappa\varphi\,\partial^\mu\bar\phi$
(see Eq.~\eqref{eq:Lgf} below), the variation
$\delta S_{\rm gf}/\delta\varphi$ is nonzero. Explicitly,
\begin{equation}
\frac{\delta S_{\rm gf}}{\delta\varphi}
=
\frac{\zeta\kappa}{\xi}\sqrt{-\bar g}\,\chi_\mu\,\partial^\mu\bar\phi,
\label{eq:Sgf-var-phi}
\end{equation}
where $\chi^\mu = h^{\mu\nu}{}_{;\nu}-\tfrac{1}{2}h^\alpha{}_{\alpha;}{}^\mu
-\zeta\kappa\varphi\,\partial^\mu\bar\phi$.
The scalar LM Lagrangian in the gauge-fixed formulation is therefore
\begin{equation}
\mathcal{L}_{{\rm LM},\sigma'}
=
\sqrt{-g}\,\sigma'\,E_\phi
+
\sigma'\,\frac{\delta S_{\rm gf}}{\delta\varphi}
=
\sqrt{-g}\,g^{\mu\nu}\partial_\mu\sigma'\,\partial_\nu\phi
+
\frac{\zeta\kappa}{\xi}\sqrt{-\bar g}\,\sigma'\,\chi_\mu\,\partial^\mu\bar\phi,
\label{eq:LLMsigma}
\end{equation}
where the first term is the invariant coupling (obtained by
integration by parts) and the second is the gauge-fixing descendant.
The gauge-fixing descendant generates additional vertices proportional
to $\zeta$; it vanishes when $\zeta=0$. With $[\phi]=1$ and
$[\sigma']=1$, no additional coupling constant is needed. The
corresponding action is
\begin{equation}
S_{{\rm LM},\sigma'}
=
\int d^4x\,\mathcal{L}_{{\rm LM},\sigma'}.
\label{eq:SLMsigma}
\end{equation}

The total action in the quantum sector is
\begin{equation}
S_{\rm tot}
=
S[g,\phi]
+ S_{\rm LM}[g,\phi,\lambda']
+ S_{{\rm LM},\sigma'}[g,\phi,\sigma']
+ S_{\rm gf}
+ S_{\rm gh},
\label{eq:Stot}
\end{equation}
where $S_{\rm gf}$ and $S_{\rm gh}$ are specified below.

The key structural property of the gauge-fixed LM action
\eqref{eq:SLMtot} is that $\lambda'_{\mu\nu}$ and $\sigma'$ both
appear \emph{linearly}. This means each vertex carries at most one
$\lambda'$ leg and at most one $\sigma'$ leg; vertices such as
$\lambda'\lambda' h$, $\sigma'\sigma'\phi$, or $\lambda'\sigma'\phi$ are
absent. As we show in Sec.~\ref{sec:obstruction}, any renormalization
of $\lambda'_{\mu\nu}$ or $\sigma'$ generates only equation-of-motion
counterterms, which vanish on shell.

\subsection{Gauge fixing}

To fix the diffeomorphism invariance of the quantum graviton sector,
we use the general background-covariant gauge-fixing term of
Ref.~\cite{Frenkel:2026zwc}, characterized by two parameters
$\xi$ and $\zeta$:
\begin{equation}
\mathcal{L}_{\rm gf}
=
-\frac{1}{2\xi}\sqrt{-\bar g}
\left(
h^{\mu\nu}{}_{;\nu}
- \frac{1}{2}h^\alpha{}_{\alpha;}{}^\mu
- \zeta\kappa\varphi\,\partial^\mu\bar\phi
\right)^2\!,
\label{eq:Lgf}
\end{equation}
where semicolons denote covariant derivatives with respect to
$\bar g_{\mu\nu}$. The parameter $\xi$ controls the standard graviton
gauge freedom, while $\zeta$ parametrizes a coupling between the
quantum scalar $\varphi$ and the gauge-fixing function through the
background scalar gradient $\partial^\mu\bar\phi$. The special cases
$\xi=\zeta=1$ and $\xi=1$, $\zeta=0$ recover the gauges of
Refs.~\cite{tHooft:1974toh} and \cite{Grisaru:1975ei}, respectively.

The corresponding Faddeev--Popov ghost Lagrangian
is~\cite{Frenkel:2026zwc}:
\begin{equation}
\mathcal{L}_{\rm gh}
=
\sqrt{-\bar g}\,\bar c^\mu
\!\left[
c_{\mu;\alpha}{}^{;\alpha}
- \bar R_{\mu\nu}c^\nu
- \zeta\kappa^2\,\partial_\mu\bar\phi\,\partial_\nu\bar\phi\,c^\nu
\right]\!,
\label{eq:Lgh}
\end{equation}
where $\bar c^\mu$ and $c^\nu$ are the vector ghost fields
(ghost indices are raised and lowered with $\bar g_{\mu\nu}$).
Since $\mathcal{L}_{\rm gf}$ depends on $h_{\mu\nu}$ and $\varphi$
through $\chi^\mu$, the gauge-fixing descendants of both LM fields
are nonzero. The tensor LM descendant
$\lambda'_{\mu\nu}\,\delta S_{\rm gf}/\delta h_{\mu\nu}$ is already
included in \eqref{eq:LLM} and accounts for the de~Donder
contribution to the graviton propagator. The scalar LM descendant
$\sigma'\,\delta S_{\rm gf}/\delta\varphi$ is included in
\eqref{eq:LLMsigma}; it vanishes when $\zeta=0$. There is no direct
LM--ghost vertex from either multiplier, since the ghost Lagrangian
\eqref{eq:Lgh} does not contain $\lambda'_{\mu\nu}$ or $\sigma'$:
\begin{equation}
V_{\lambda'\bar c\, c} = 0,
\qquad
V_{\sigma'\bar c\, c} = 0.
\label{eq:no-lambda-ghost}
\end{equation}

\subsection{Propagators}

Around a flat background ($\bar g_{\mu\nu} = \eta_{\mu\nu}$), the
quadratic sector in the $(h_{\mu\nu},\lambda'_{\mu\nu})$ subsystem
takes the form
\begin{equation}
\mathcal{L}^{(2)}_{h\lambda'}
=
\frac{1}{2}h^{\mu\nu}O_{\mu\nu,\rho\sigma}h^{\rho\sigma}
+
\lambda'^{\mu\nu}O_{\mu\nu,\rho\sigma}h^{\rho\sigma},
\label{eq:quadratic}
\end{equation}
with $O_{\mu\nu,\rho\sigma}(k) = -k^2 P_{\mu\nu,\rho\sigma}$ and
\begin{equation}
P_{\mu\nu,\rho\sigma}
=
\frac{1}{2}
\!\left(
\eta_{\mu\rho}\eta_{\nu\sigma}
+\eta_{\mu\sigma}\eta_{\nu\rho}
-\eta_{\mu\nu}\eta_{\rho\sigma}
\right)\!.
\label{eq:P-tensor}
\end{equation}
This is the de~Donder gauge ($\xi=1$) form; the general
$(\xi,\zeta)$-dependent propagator tensor is given in
Eq.~(A12) of Ref.~\cite{Frenkel:2026zwc}.
The corresponding kernel and its inverse are
\begin{equation}
K
=
\begin{pmatrix}
O & O \\
O & 0
\end{pmatrix},
\quad
K^{-1}
=
\begin{pmatrix}
0 & O^{-1} \\
O^{-1} & -O^{-1}
\end{pmatrix},
\label{eq:kernel}
\end{equation}
giving the quantum propagators
\begin{align}
\langle h_{\mu\nu}h_{\rho\sigma}\rangle_q(k)
&= 0,
\label{eq:hh-zero}
\\
\langle h_{\mu\nu}\lambda'_{\rho\sigma}\rangle_q(k)
=
\langle\lambda'_{\mu\nu}h_{\rho\sigma}\rangle_q(k)
&=
-\frac{i\,P_{\mu\nu,\rho\sigma}}{k^2+i\varepsilon},
\label{eq:hl-prop}
\\
\langle\lambda'_{\mu\nu}\lambda'_{\rho\sigma}\rangle_q(k)
&=
+\frac{i\,P_{\mu\nu,\rho\sigma}}{k^2+i\varepsilon}.
\label{eq:ll-prop}
\end{align}
The vanishing of \(\langle hh\rangle_q\) is the central structural
feature of the LM formalism: internal gravitational-sector lines are
replaced by the nonvanishing \(h\lambda'\), \(\lambda' h\), and
\(\lambda'\lambda'\) propagators. Note the
relation
\begin{equation}
\langle\lambda'\lambda'\rangle_q
=
-\langle h\lambda'\rangle_q
=
-\langle\lambda' h\rangle_q,
\label{eq:propagator-relation}
\end{equation}
which plays a key role in the diagrammatic analysis of
Sec.~\ref{sec:CT}; the Feynman rules are collected in
Appendix~\ref{app:diagrams}.

The quantum scalar sector is now modified by the presence of the
scalar multiplier $\sigma'$. Around a flat background, the quadratic
Lagrangian in $(\varphi,\sigma')$ is
\begin{equation}
\mathcal{L}^{(2)}_{\varphi\sigma'}
=
\frac{1}{2}\,\partial_\mu\varphi\,\partial^\mu\varphi
+
\partial_\mu\sigma'\,\partial^\mu\varphi ,
\label{eq:quadratic-scalar}
\end{equation}
the first term from $S_\phi$ and the second from
$S_{{\rm LM},\sigma'}$, Eq.~\eqref{eq:LLMsigma}. With
$O_\phi(k)=+k^2$ this has exactly the block structure of the
gravitational sector,
\begin{equation}
K_{\varphi\sigma'}
=
\begin{pmatrix}
O_\phi & O_\phi \\
O_\phi & 0
\end{pmatrix},
\quad
K_{\varphi\sigma'}^{-1}
=
\begin{pmatrix}
0 & O_\phi^{-1} \\
O_\phi^{-1} & -O_\phi^{-1}
\end{pmatrix},
\label{eq:kernel-scalar}
\end{equation}
giving the quantum matter propagators
\begin{align}
\langle\varphi\varphi\rangle_q(k)
&= 0,
\label{eq:phiphi-zero}
\\
\langle\varphi\sigma'\rangle_q(k)
=
\langle\sigma'\varphi\rangle_q(k)
&=
\frac{i}{k^2+i\varepsilon},
\label{eq:phisigma-prop}
\\
\langle\sigma'\sigma'\rangle_q(k)
&=
-\frac{i}{k^2+i\varepsilon}.
\label{eq:sigmasigma-prop}
\end{align}
The vanishing of $\langle\varphi\varphi\rangle_q$ is the matter
analogue of $\langle hh\rangle_q=0$: internal quantum-scalar lines
are replaced by the nonvanishing $\varphi\sigma'$, $\sigma'\varphi$,
and $\sigma'\sigma'$ propagators, which satisfy
\begin{equation}
\langle\sigma'\sigma'\rangle_q
=
-\langle\varphi\sigma'\rangle_q
=
-\langle\sigma'\varphi\rangle_q ,
\label{eq:propagator-relation-scalar}
\end{equation}
in exact parallel with Eq.~\eqref{eq:propagator-relation}. Around
flat space with vanishing background scalar gradient, the
$(\varphi,\sigma')$ sector decouples from the $(h,\lambda')$ sector at
quadratic order, so there is no $\langle\sigma' h\rangle$ or
$\langle\sigma'\lambda'\rangle$ mixing in this limit. In a general
background with $\partial_\mu\bar\phi\neq0$, however, both the
invariant coupling, through the $\sigma'\bar\phi h$ vertex of
Eq.~\eqref{eq:cubic-scalar-vertices}, and the gauge-fixing descendant
$\sigma'\,\delta S_{\rm gf}/\delta\varphi$ produce $h$--$\sigma'$ mixing; this mixing is part of the full
coupled field-space operator $O_{AB}$. The diagrammatic
forms of all propagators are collected in Appendix~\ref{app:diagrams}.

The ghost propagator is the standard one,
\begin{equation}
\langle c^\mu\bar c_\nu\rangle_q(k)
=
\frac{i\,\delta^\mu{}_\nu}{k^2+i\varepsilon}.
\label{eq:ghost-prop}
\end{equation}

The block structure of the gauge-fixed LM quadratic matrix
\eqref{eq:LM-block} implies, via a standard block-determinant
identity, that the LM path integral over the full field space
$Q^A=(h_{\mu\nu},\varphi)$ produces a squared functional determinant
relative to the ordinary gauge-fixed formalism:
\begin{equation}
\det K
=
\det
\begin{pmatrix} O & O \\ O & 0 \end{pmatrix}
=
(-1)^{N/2}(\det O)^2,
\label{eq:det-doubling}
\end{equation}
where $O_{AB}$ is the full gauge-fixed Hessian of $S_{\rm inv}+S_{\rm gf}$
in the coupled $(h_{\mu\nu},\varphi)$ field space, and $N$ is the total number of components in the
full doubled field space (equivalently, twice the dimension of the
ordinary gauge-fixed Hessian block). The determinant doubling is
independent of the LM fields and holds for the \emph{coupled}
operator $O_{AB}$, not separately for the graviton and scalar
sectors; in a nontrivial scalar background, $O_{AB}$ generally
mixes $h_{\mu\nu}$ and $\varphi$.

Lee--Yang ghost fields, denoted here by $\eta,\bar\eta$ (and by
$\eta_s,\bar\eta_s$ in the scalar sector) to avoid confusion with the
gauge-fixing function $\chi^\mu$, are
introduced to compensate this doubling. The Gaussian integral over
$(Q^A,\Lambda_A)$ gives $(\det K)^{-1/2}\propto(\det O)^{-1}$, whereas
the ordinary theory gives $(\det O)^{-1/2}$; the Lee--Yang sector
therefore supplies the compensating factor $(\det O)^{+1/2}$. Their action is built from the
same gauge-fixed Hessian $O_{AB}$ and therefore does not depend on
the LM fields $\lambda'_{\mu\nu}$ or $\sigma'$. As a result, there
are no vertices between the LM fields and the Lee--Yang ghosts:
\begin{equation}
V^{\rm LY}_{\lambda'\bar\eta\eta} = 0,
\qquad
V^{\rm LY}_{\sigma'\bar\eta_s\eta_s} = 0.
\label{eq:LY-vertices}
\end{equation}
After this cancellation, the LM path integral is equivalent to the
ordinary gauge-fixed path integral of Ref.~\cite{Frenkel:2026zwc},
and the one-loop counterterm coefficients $\mathcal{C}_i(\xi,\zeta)$
may be imported from that reference. At the level of the full functional determinant, the LM system gives
a doubling of the ordinary one-loop contribution, which is cancelled
by the Lee--Yang determinant. Diagrammatically, this doubling is
reflected in the mixed-propagator representation of the ordinary
one-loop topologies; however, individual diagrams should be
constructed from the full gauge-fixed LM vertex set rather than by
an independent line-by-line replacement. The net effect, after the
Lee--Yang cancellation, is that the LM and ordinary one-loop
counterterm Lagrangians agree, as discussed in Sec.~\ref{sec:CT}.

\subsection{Vertices}

For the invariant scalar kinetic part, the tensor and scalar LM
vertices are obtained from the corresponding ordinary vertices by
replacing one quantum field by its LM partner. Thus
\begin{equation}
V^{{\rm inv},\mu\nu}_{\lambda'\phi\phi}(p,q)
=
V^{{\rm inv},\mu\nu}_{h\phi\phi}(p,q)
=
-\frac{i\kappa}{2}\,\tau^{\mu\nu}(p,q),
\label{eq:lambda-phi-phi}
\end{equation}
where
\begin{equation}
\tau^{\mu\nu}(p,q)
=
p^\mu q^\nu + p^\nu q^\mu
- \eta^{\mu\nu}\,p\cdot q,
\label{eq:tau}
\end{equation}
with \(p\) and \(q\) the two incoming scalar momenta.
In the full gauge-fixed theory these identities must be supplemented
by the gauge-fixing descendant vertices generated by
\(\lambda'_{\mu\nu}\delta S_{\rm gf}/\delta h_{\mu\nu}\) and
\(\sigma'\,\delta S_{\rm gf}/\delta\varphi\).

Similarly, the LM gravitational vertices are obtained from the
ordinary Einstein--Hilbert vertices by replacing one graviton leg
with a $\lambda'$ leg. Since $\lambda'_{\mu\nu}$ is a distinct field,
the resulting vertex has no symmetry between the $\lambda'$ leg and
the graviton legs; only the graviton legs among themselves are
symmetrized:
\begin{equation}
V^{(g)}_{\lambda' H_1\cdots H_n}
=
V^{\rm EH}_{H H_1\cdots H_n}\Big|_{H\to\lambda'},
\quad
H_i \in \{\bar h,\, h\},
\label{eq:lambda-grav-rule}
\end{equation}
with at most one $\lambda'$ leg per vertex since the action is linear
in $\lambda'_{\mu\nu}$.

The scalar multiplier $\sigma'$ generates vertices in the same way.
For the invariant scalar LM coupling, the graviton--$\sigma'$--scalar
vertices satisfy the matter LM identity
\begin{equation}
V^{{\rm inv},\mu\nu}_{H\sigma'\phi}(p,q)
=
V^{{\rm inv},\mu\nu}_{H\phi\phi}(p,q)
=
-\frac{i\kappa}{2}\,\tau^{\mu\nu}(p,q),
\qquad
H\in\{\bar h,\,h\},
\label{eq:sigma-vertex}
\end{equation}
with $p$ the $\sigma'$ momentum and $q$ the $\phi$ momentum. This is
the matter analogue of Eq.~\eqref{eq:lambda-phi-phi}. Since $\sigma'$
appears linearly, every such vertex has exactly one $\sigma'$ leg, and
there are no $\sigma'\sigma'$ or mixed $\lambda'\sigma'$ vertices.

The invariant vertex identities \eqref{eq:lambda-phi-phi} and
\eqref{eq:sigma-vertex}, together with the gauge-fixing descendant
vertices and the propagator relations~\eqref{eq:propagator-relation}
and \eqref{eq:propagator-relation-scalar}, determine the allowed LM
diagrams. The diagrammatic Feynman rules are given in
Appendix~\ref{app:diagrams}, and the diagrammatic analysis is
discussed in Sec.~\ref{sec:CT}.

\section{One-loop renormalization in the LM formulation}
\label{sec:CT}

\subsection{One-loop diagrams with mixed internal lines}
\label{sec:diagrams}

In the LM formulation the quantum graviton propagator vanishes,
$\langle hh\rangle_q=0$, as does the quantum scalar propagator,
$\langle\varphi\varphi\rangle_q=0$ (Sec.~\ref{sec:formulation}).
The one-loop diagrams are therefore built entirely from the nonvanishing
mixed propagators: $\langle h\lambda'\rangle_q$,
$\langle\lambda' h\rangle_q$, $\langle\lambda'\lambda'\rangle_q$ in the
gravitational sector, and $\langle\varphi\sigma'\rangle_q$,
$\langle\sigma'\varphi\rangle_q$, $\langle\sigma'\sigma'\rangle_q$ in the
scalar sector (Appendix~\ref{app:diagrams}).

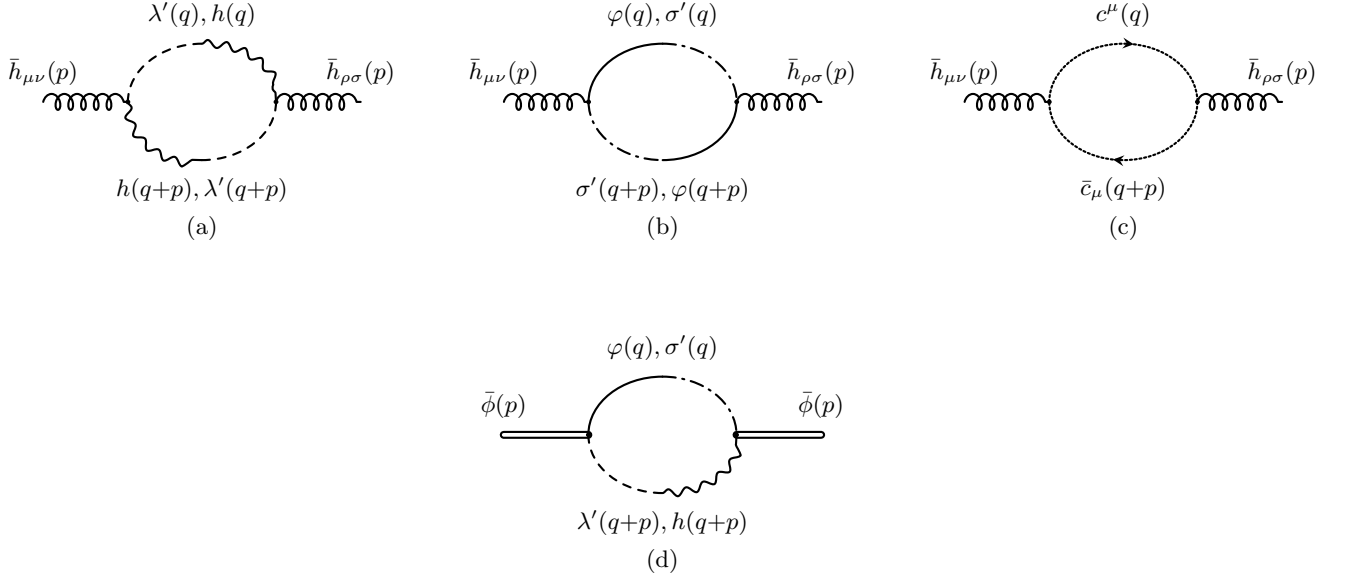
\begin{figure}[ht]
\begin{center}
\begin{tikzpicture}[baseline=-0.2ex,scale=0.70,line cap=round,line join=round]
  \SelfEnergyCoords
  \draw[bgrav](Lext)--(L);
  \draw[bgrav](R)--(Rext);
  \UpperLeft{qlm}\UpperRight{qgrav}
  \LowerLeft{qgrav}\LowerRight{qlm}
  \Vertices
  \node[above] at (-3.0,0.15){\small$\bar h_{\mu\nu}(p)$};
  \node[above] at ( 3.0,0.15){\small$\bar h_{\rho\sigma}(p)$};
  \node[above] at (0,1.25){\small$\lambda'(q),h(q)$};
  \node[below] at (0,-1.25){\small$h(q{+}p),\lambda'(q{+}p)$};
  \node[below] at (0,-2.0){\small(a)};
\end{tikzpicture}
\qquad
\begin{tikzpicture}[baseline=-0.2ex,scale=0.70,line cap=round,line join=round]
  \SelfEnergyCoords
  \draw[bgrav](Lext)--(L);
  \draw[bgrav](R)--(Rext);
  \UpperLeft{qscalar}\UpperRight{qslm}
  \LowerLeft{qslm}\LowerRight{qscalar}
  \Vertices
  \node[above] at (-3.0,0.15){\small$\bar h_{\mu\nu}(p)$};
  \node[above] at ( 3.0,0.15){\small$\bar h_{\rho\sigma}(p)$};
  \node[above] at (0,1.25){\small$\varphi(q),\sigma'(q)$};
  \node[below] at (0,-1.25){\small$\sigma'(q{+}p),\varphi(q{+}p)$};
  \node[below] at (0,-2.0){\small(b)};
\end{tikzpicture}
\qquad
\begin{tikzpicture}[baseline=-0.2ex,scale=0.70,line cap=round,line join=round]
  \SelfEnergyCoords
  \draw[bgrav](Lext)--(L);
  \draw[bgrav](R)--(Rext);
  \draw[ghost,postaction={decorate,decoration={markings,
    mark=at position 0.55 with{\arrow[>=stealth]{>}}}}]
    (L) arc[start angle=180,end angle=0,x radius=1.4,y radius=1.1];
  \draw[ghost,postaction={decorate,decoration={markings,
    mark=at position 0.55 with{\arrow[>=stealth]{>}}}}]
    (R) arc[start angle=0,end angle=-180,x radius=1.4,y radius=1.1];
  \Vertices
  \node[above] at (-3.0,0.15){\small$\bar h_{\mu\nu}(p)$};
  \node[above] at ( 3.0,0.15){\small$\bar h_{\rho\sigma}(p)$};
  \node[above] at (0,1.25){\small$c^\mu(q)$};
  \node[below] at (0,-1.25){\small$\bar c_\mu(q{+}p)$};
  \node[below] at (0,-2.0){\small(c)};
\end{tikzpicture}
\\[10mm]
\begin{tikzpicture}[baseline=-0.2ex,scale=0.70,line cap=round,line join=round]
  \SelfEnergyCoords
  \draw[bscalar](Lext)--(L);
  \draw[bscalar](R)--(Rext);
  \UpperLeft{qscalar}\UpperRight{qslm}
  \LowerLeft{qlm}\LowerRight{qgrav}
  \Vertices
  \node[above] at (-3.0,0.15){\small$\bar\phi(p)$};
  \node[above] at ( 3.0,0.15){\small$\bar\phi(p)$};
  \node[above] at (0,1.25){\small$\varphi(q),\sigma'(q)$};
  \node[below] at (0,-1.25){\small$\lambda'(q{+}p),h(q{+}p)$};
  \node[below] at (0,-2.0){\small(d)};
\end{tikzpicture}
\end{center}
\caption{One-loop contributions to $\langle\bar h_{\mu\nu}\bar h_{\rho\sigma}\rangle$
[diagrams (a), (b), (c)] and to $\langle\bar\phi\,\bar\phi\rangle$ [diagram (d)].
In (a) and (d) the internal lines carry mixed propagators; (b) has
mixed matter propagators; (c) has ordinary ghost propagators.}
\label{fig:2pt}
\end{figure}

The one-loop contributions to the background graviton self-energy
$\langle\bar h_{\mu\nu}\bar h_{\rho\sigma}\rangle$
and the background scalar self-energy $\langle\bar\phi\,\bar\phi\rangle$
are shown in Fig.~\ref{fig:2pt}. In each diagram every internal loop
line carries a mixed propagator; in the gravitational loops the lines
are $\langle h\lambda'\rangle$ (or $\langle\lambda' h\rangle$,
$\langle\lambda'\lambda'\rangle$), and in the matter loop and the scalar
self-energy the lines are $\langle\varphi\sigma'\rangle$ (or
$\langle\sigma'\varphi\rangle$, $\langle\sigma'\sigma'\rangle$).

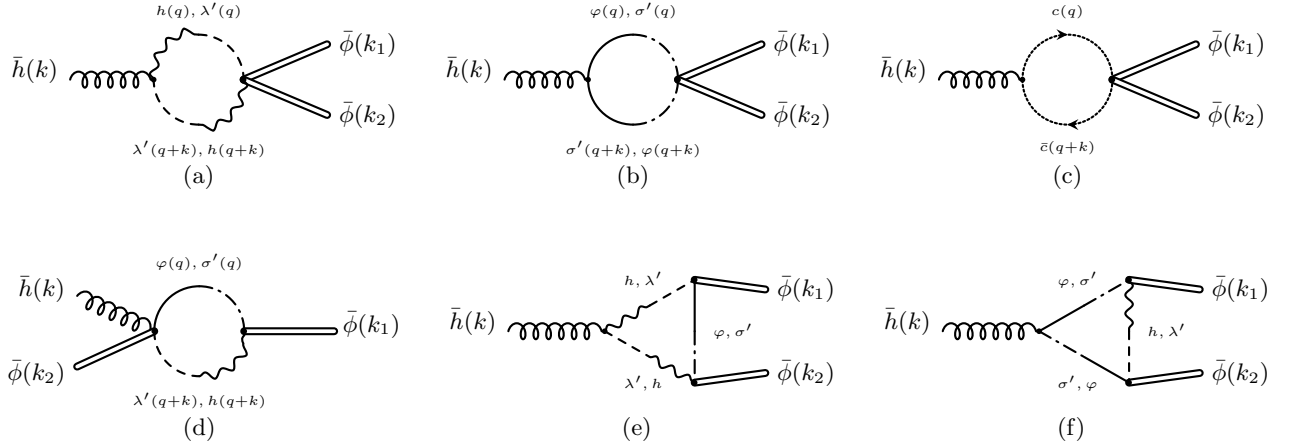
\begin{figure}[ht]
\begin{center}
\begin{tikzpicture}[baseline=0ex,scale=0.85,line cap=round,line join=round]
  \coordinate (L) at (-0.7,0); \coordinate (R) at (0.7,0);
  \coordinate (M) at (0,0.7);  \coordinate (Mb) at (0,-0.7);
  \coordinate (P1) at (-2.0,0);
  \coordinate (P2) at (2.0,0.55); \coordinate (P3) at (2.0,-0.55);
  \draw[bgrav](P1)--(L); \draw[bscalar](R)--(P2); \draw[bscalar](R)--(P3);
  \draw[qgrav](L) arc[start angle=180,end angle=90,x radius=0.7,y radius=0.7];
  \draw[qlm](M) arc[start angle=90,end angle=0,x radius=0.7,y radius=0.7];
  \draw[qgrav](R) arc[start angle=0,end angle=-90,x radius=0.7,y radius=0.7];
  \draw[qlm](Mb) arc[start angle=-90,end angle=-180,x radius=0.7,y radius=0.7];
  \vertexdot{L}\vertexdot{R}
  \node[left]  at (-2.05,0.15){\small$\bar h(k)$};
  \node[right] at (2.05,0.55){\small$\bar\phi(k_1)$};
  \node[right] at (2.05,-0.55){\small$\bar\phi(k_2)$};
  \node[above] at (0,0.82){\tiny$h(q),\lambda'(q)$};
  \node[below] at (0,-0.82){\tiny$\lambda'(q{+}k),h(q{+}k)$};
  \node[below] at (0,-1.2){\small(a)};
\end{tikzpicture}
\quad
\begin{tikzpicture}[baseline=0ex,scale=0.85,line cap=round,line join=round]
  \coordinate (L) at (-0.7,0); \coordinate (R) at (0.7,0);
  \coordinate (M) at (0,0.7);  \coordinate (Mb) at (0,-0.7);
  \coordinate (P1) at (-2.0,0);
  \coordinate (P2) at (2.0,0.55); \coordinate (P3) at (2.0,-0.55);
  \draw[bgrav](P1)--(L); \draw[bscalar](R)--(P2); \draw[bscalar](R)--(P3);
  \draw[qscalar](L) arc[start angle=180,end angle=90,x radius=0.7,y radius=0.7];
  \draw[qslm](M) arc[start angle=90,end angle=0,x radius=0.7,y radius=0.7];
  \draw[qslm](R) arc[start angle=0,end angle=-90,x radius=0.7,y radius=0.7];
  \draw[qscalar](Mb) arc[start angle=-90,end angle=-180,x radius=0.7,y radius=0.7];
  \vertexdot{L}\vertexdot{R}
  \node[left]  at (-2.05,0.15){\small$\bar h(k)$};
  \node[right] at (2.05,0.55){\small$\bar\phi(k_1)$};
  \node[right] at (2.05,-0.55){\small$\bar\phi(k_2)$};
  \node[above] at (0,0.82){\tiny$\varphi(q),\sigma'(q)$};
  \node[below] at (0,-0.82){\tiny$\sigma'(q{+}k),\varphi(q{+}k)$};
  \node[below] at (0,-1.2){\small(b)};
\end{tikzpicture}
\quad
\begin{tikzpicture}[baseline=0ex,scale=0.85,line cap=round,line join=round]
  \coordinate (L) at (-0.7,0); \coordinate (R) at (0.7,0);
  \coordinate (P1) at (-2.0,0);
  \coordinate (P2) at (2.0,0.55); \coordinate (P3) at (2.0,-0.55);
  \draw[bgrav](P1)--(L); \draw[bscalar](R)--(P2); \draw[bscalar](R)--(P3);
  \draw[ghost,postaction={decorate,decoration={markings,
    mark=at position 0.5 with{\arrow[>=stealth]{>}}}}]
    (L) arc[start angle=180,end angle=0,x radius=0.7,y radius=0.7];
  \draw[ghost,postaction={decorate,decoration={markings,
    mark=at position 0.5 with{\arrow[>=stealth]{>}}}}]
    (R) arc[start angle=0,end angle=-180,x radius=0.7,y radius=0.7];
  \vertexdot{L}\vertexdot{R}
  \node[left] at (-2.05,0.15){\small$\bar h(k)$};
  \node[right] at (2.05,0.55){\small$\bar\phi(k_1)$};
  \node[right] at (2.05,-0.55){\small$\bar\phi(k_2)$};
  \node[above] at (0,0.82){\tiny$c(q)$};
  \node[below] at (0,-0.82){\tiny$\bar c(q{+}k)$};
  \node[below] at (0,-1.2){\small(c)};
\end{tikzpicture}
\\[6mm]
\begin{tikzpicture}[baseline=0ex,scale=0.85,line cap=round,line join=round]
  \coordinate (L)  at (-0.7,0); \coordinate (R)  at (0.7,0);
  \coordinate (P1) at (-1.9,0.55); \coordinate (P2) at (-1.9,-0.55);
  \coordinate (P3) at (2.1,0);
  \draw[bgrav](P1)--(L); \draw[bscalar](P2)--(L); \draw[bscalar](R)--(P3);
  \draw[qscalar](L) arc[start angle=180,end angle=90,x radius=0.7,y radius=0.7];
  \draw[qslm]   (0,0.7) arc[start angle=90,end angle=0,x radius=0.7,y radius=0.7];
  \draw[qlm](L) arc[start angle=180,end angle=270,x radius=0.7,y radius=0.7];
  \draw[qgrav](0,-0.7) arc[start angle=270,end angle=360,x radius=0.7,y radius=0.7];
  \vertexdot{L}\vertexdot{R}
  \node[left]  at (-1.95,0.65){\small$\bar h(k)$};
  \node[left]  at (-1.95,-0.65){\small$\bar\phi(k_2)$};
  \node[right] at (2.1,0.1){\small$\bar\phi(k_1)$};
  \node[above] at (0,0.82){\tiny$\varphi(q),\sigma'(q)$};
  \node[below] at (0,-0.82){\tiny$\lambda'(q{+}k),h(q{+}k)$};
  \node[below] at (0,-1.2){\small(d)};
\end{tikzpicture}
\quad
\begin{tikzpicture}[baseline=0ex,scale=0.85,line cap=round,line join=round]
  \coordinate (VL) at (-1.5,0); \coordinate (V1) at (0,0);
  \coordinate (V2) at (1.4,0.8); \coordinate (V3) at (1.4,-0.8);
  \coordinate (P1) at (2.5,0.65); \coordinate (P2) at (2.5,-0.65);
  \draw[bgrav](VL)--(V1);
  \draw[bscalar](V2)--(P1); \draw[bscalar](V3)--(P2);
  \mixedWDc{V1}{V2}
  \mixedDWc{V1}{V3}
  \mixedSscalarc{V2}{V3}
  \vertexdot{V1}\vertexdot{V2}\vertexdot{V3}
  \node[left]  at (-1.55,0.15){\small$\bar h(k)$};
  \node[right] at (2.55,0.65){\small$\bar\phi(k_1)$};
  \node[right] at (2.55,-0.65){\small$\bar\phi(k_2)$};
  \node[above] at (0.6,0.55){\tiny$h,\lambda'$};
  \node[below] at (0.6,-0.55){\tiny$\lambda',h$};
  \node[right] at (1.55,0.0){\tiny$\varphi,\sigma'$};
  \node[below] at (0.5,-1.2){\small(e)};
\end{tikzpicture}
\quad
\begin{tikzpicture}[baseline=0ex,scale=0.85,line cap=round,line join=round]
  \coordinate (VL) at (-1.5,0); \coordinate (V1) at (0,0);
  \coordinate (V2) at (1.4,0.8); \coordinate (V3) at (1.4,-0.8);
  \coordinate (P1) at (2.5,0.65); \coordinate (P2) at (2.5,-0.65);
  \draw[bgrav](VL)--(V1);
  \draw[bscalar](V2)--(P1); \draw[bscalar](V3)--(P2);
  \mixedSscalarc{V1}{V2}
  \mixedSscalarRevc{V1}{V3}
  \mixedWDc{V2}{V3}
  \vertexdot{V1}\vertexdot{V2}\vertexdot{V3}
  \node[left]  at (-1.55,0.15){\small$\bar h(k)$};
  \node[right] at (2.55,0.65){\small$\bar\phi(k_1)$};
  \node[right] at (2.55,-0.65){\small$\bar\phi(k_2)$};
  \node[above] at (0.6,0.55){\tiny$\varphi,\sigma'$};
  \node[below] at (0.6,-0.55){\tiny$\sigma',\varphi$};
  \node[right] at (1.55,0.0){\tiny$h,\lambda'$};
  \node[below] at (0.5,-1.2){\small(f)};
\end{tikzpicture}
\end{center}
\caption{One-loop contributions to
$\langle\bar h_{\mu\nu}(k)\bar\phi(k_1)\bar\phi(k_2)\rangle$,
corresponding to diagrams (a)--(f) of Fig.~2 of Ref.~\cite{Frenkel:2026zwc}
(permutations omitted). Internal lines carry mixed propagators,
except for the ghost loop (c).}
\label{fig:3pt}
\end{figure}

The one-loop contributions to the three-point function
$\langle\bar h_{\mu\nu}\bar\phi\,\bar\phi\rangle$ are shown in
Fig.~\ref{fig:3pt}.

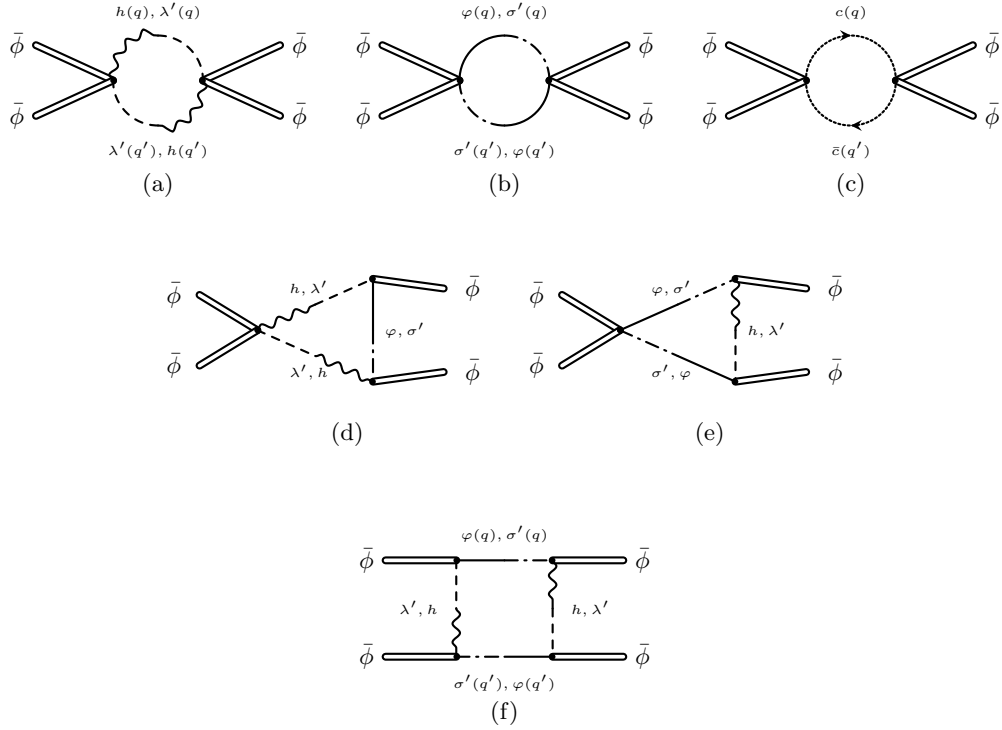
\begin{figure}[ht]
\begin{center}
\begin{tikzpicture}[baseline=0ex,scale=0.85,line cap=round,line join=round]
  \coordinate (L) at (-0.7,0); \coordinate (R) at (0.7,0);
  \coordinate (M) at (0,0.7);  \coordinate (Mb) at (0,-0.7);
  \coordinate (P1) at (-1.9,0.55); \coordinate (P2) at (-1.9,-0.55);
  \coordinate (P3) at (1.9,0.55);  \coordinate (P4) at (1.9,-0.55);
  \draw[bscalar](P1)--(L); \draw[bscalar](P2)--(L);
  \draw[bscalar](R)--(P3); \draw[bscalar](R)--(P4);
  \draw[qgrav](L) arc[start angle=180,end angle=90,x radius=0.7,y radius=0.7];
  \draw[qlm](M) arc[start angle=90,end angle=0,x radius=0.7,y radius=0.7];
  \draw[qgrav](R) arc[start angle=0,end angle=-90,x radius=0.7,y radius=0.7];
  \draw[qlm](Mb) arc[start angle=-90,end angle=-180,x radius=0.7,y radius=0.7];
  \vertexdot{L}\vertexdot{R}
  \node[left] at (-1.95,0.55){\small$\bar\phi$};
  \node[left] at (-1.95,-0.55){\small$\bar\phi$};
  \node[right] at (1.95,0.55){\small$\bar\phi$};
  \node[right] at (1.95,-0.55){\small$\bar\phi$};
  \node[above] at (0,0.82){\tiny$h(q),\lambda'(q)$};
  \node[below] at (0,-0.82){\tiny$\lambda'(q'),h(q')$};
  \node[below] at (0,-1.3){\small(a)};
\end{tikzpicture}
\quad
\begin{tikzpicture}[baseline=0ex,scale=0.85,line cap=round,line join=round]
  \coordinate (L) at (-0.7,0); \coordinate (R) at (0.7,0);
  \coordinate (M) at (0,0.7);  \coordinate (Mb) at (0,-0.7);
  \coordinate (P1) at (-1.9,0.55); \coordinate (P2) at (-1.9,-0.55);
  \coordinate (P3) at (1.9,0.55);  \coordinate (P4) at (1.9,-0.55);
  \draw[bscalar](P1)--(L); \draw[bscalar](P2)--(L);
  \draw[bscalar](R)--(P3); \draw[bscalar](R)--(P4);
  \draw[qscalar](L) arc[start angle=180,end angle=90,x radius=0.7,y radius=0.7];
  \draw[qslm](M) arc[start angle=90,end angle=0,x radius=0.7,y radius=0.7];
  \draw[qscalar](R) arc[start angle=0,end angle=-90,x radius=0.7,y radius=0.7];
  \draw[qslm](Mb) arc[start angle=-90,end angle=-180,x radius=0.7,y radius=0.7];
  \vertexdot{L}\vertexdot{R}
  \node[left] at (-1.95,0.55){\small$\bar\phi$};
  \node[left] at (-1.95,-0.55){\small$\bar\phi$};
  \node[right] at (1.95,0.55){\small$\bar\phi$};
  \node[right] at (1.95,-0.55){\small$\bar\phi$};
  \node[above] at (0,0.82){\tiny$\varphi(q),\sigma'(q)$};
  \node[below] at (0,-0.82){\tiny$\sigma'(q'),\varphi(q')$};
  \node[below] at (0,-1.3){\small(b)};
\end{tikzpicture}
\quad
\begin{tikzpicture}[baseline=0ex,scale=0.85,line cap=round,line join=round]
  \coordinate (L) at (-0.7,0); \coordinate (R) at (0.7,0);
  \coordinate (P1) at (-1.9,0.55); \coordinate (P2) at (-1.9,-0.55);
  \coordinate (P3) at (1.9,0.55);  \coordinate (P4) at (1.9,-0.55);
  \draw[bscalar](P1)--(L); \draw[bscalar](P2)--(L);
  \draw[bscalar](R)--(P3); \draw[bscalar](R)--(P4);
  \draw[ghost,postaction={decorate,decoration={markings,
    mark=at position 0.5 with{\arrow[>=stealth]{>}}}}]
    (L) arc[start angle=180,end angle=0,x radius=0.7,y radius=0.7];
  \draw[ghost,postaction={decorate,decoration={markings,
    mark=at position 0.5 with{\arrow[>=stealth]{>}}}}]
    (R) arc[start angle=0,end angle=-180,x radius=0.7,y radius=0.7];
  \vertexdot{L}\vertexdot{R}
  \node[left] at (-1.95,0.55){\small$\bar\phi$};
  \node[left] at (-1.95,-0.55){\small$\bar\phi$};
  \node[right] at (1.95,0.55){\small$\bar\phi$};
  \node[right] at (1.95,-0.55){\small$\bar\phi$};
  \node[above] at (0,0.82){\tiny$c(q)$};
  \node[below] at (0,-0.82){\tiny$\bar c(q')$};
  \node[below] at (0,-1.3){\small(c)};
\end{tikzpicture}
\\[8mm]
\begin{tikzpicture}[baseline=0ex,scale=0.85,line cap=round,line join=round]
  \coordinate (VL) at (-1.4,0.0); 
  \coordinate (V1) at (-1.4,0);
  \coordinate (V2) at (0.4,0.8); 
  \coordinate (V3) at (0.4,-0.8);
  \coordinate (P1) at (-2.3,0.55); 
  \coordinate (P2) at (-2.3,-0.55);
  \coordinate (P3) at (1.5,0.65); 
  \coordinate (P4) at (1.5,-0.65);
  \draw[bscalar](P1)--(VL); 
  \draw[bscalar](P2)--(VL);
  \draw[bscalar](V2)--(P3); \draw[bscalar](V3)--(P4);
  \mixedWDc{V1}{V2}
  \mixedDWc{V1}{V3}
  \mixedSscalarc{V2}{V3}
  \vertexdot{VL}\vertexdot{V1}\vertexdot{V2}\vertexdot{V3}
  \node[left] at (-2.45,0.55){\small$\bar\phi$};
  \node[left] at (-2.45,-0.55){\small$\bar\phi$};
  \node[right] at (1.7,0.65){\small$\bar\phi$};
  \node[right] at (1.7,-0.65){\small$\bar\phi$};
  \node[above] at (-0.6,0.4){\tiny$h,\lambda'$};
  \node[below] at (-0.6,-0.4){\tiny$\lambda',h$};
  \node[right] at (0.45,0){\tiny$\varphi,\sigma'$};
  \node[below] at (0,-1.3){\small(d)};
\end{tikzpicture}
\quad
\begin{tikzpicture}[baseline=0ex,scale=0.85,line cap=round,line join=round]
  \coordinate (VL) at (-1.4,0.0);
  \coordinate (V1) at (-1.4,0);
  \coordinate (V2) at (0.4,0.8);
  \coordinate (V3) at (0.4,-0.8);
  \coordinate (P1) at (-2.3,0.55);
  \coordinate (P2) at (-2.3,-0.55);
  \coordinate (P3) at (1.5,0.65);
  \coordinate (P4) at (1.5,-0.65);
  \draw[bscalar](P1)--(VL);
  \draw[bscalar](P2)--(VL);
  \draw[bscalar](V2)--(P3); \draw[bscalar](V3)--(P4);
  \mixedSscalarc{V1}{V2}
  \mixedSscalarRevc{V1}{V3}
  \mixedWDc{V2}{V3}
  \vertexdot{VL}\vertexdot{V1}\vertexdot{V2}\vertexdot{V3}
  \node[left] at (-2.45,0.55){\small$\bar\phi$};
  \node[left] at (-2.45,-0.55){\small$\bar\phi$};
  \node[right] at (1.7,0.65){\small$\bar\phi$};
  \node[right] at (1.7,-0.65){\small$\bar\phi$};
  \node[above] at (-0.6,0.4){\tiny$\varphi,\sigma'$};
  \node[below] at (-0.6,-0.4){\tiny$\sigma',\varphi$};
  \node[right] at (0.45,0){\tiny$h,\lambda'$};
  \node[below] at (0,-1.3){\small(e)};
\end{tikzpicture}
\\[8mm]
\begin{tikzpicture}[baseline=0ex,scale=0.85,line cap=round,line join=round]
  \coordinate (TL) at (-0.75,0.75); \coordinate (TR) at (0.75,0.75);
  \coordinate (BL) at (-0.75,-0.75);\coordinate (BR) at (0.75,-0.75);
  \coordinate (P1) at (-1.85,0.75);\coordinate (P2) at (-1.85,-0.75);
  \coordinate (P3) at (1.85,0.75); \coordinate (P4) at (1.85,-0.75);
  \draw[bscalar](P1)--(TL); \draw[bscalar](P2)--(BL);
  \draw[bscalar](TR)--(P3); \draw[bscalar](BR)--(P4);
  \mixedSscalarc{TL}{TR}
  \mixedSscalarRevc{BL}{BR}
  \mixedDWc{TL}{BL}
  \mixedWDc{TR}{BR}
  \vertexdot{TL}\vertexdot{TR}\vertexdot{BL}\vertexdot{BR}
  \node[left] at (-1.9,0.75){\small$\bar\phi$};
  \node[left] at (-1.9,-0.75){\small$\bar\phi$};
  \node[right] at (1.9,0.75){\small$\bar\phi$};
  \node[right] at (1.9,-0.75){\small$\bar\phi$};
  \node[above] at (0,0.9){\tiny$\varphi(q),\sigma'(q)$};
  \node[below] at (0,-0.9){\tiny$\sigma'(q'),\varphi(q')$};
  \node[left]  at (-0.9,0){\tiny$\lambda',h$};
  \node[right] at (0.9,0){\tiny$h,\lambda'$};
  \node[below] at (0,-1.3){\small(f)};
\end{tikzpicture}
\end{center}
\caption{One-loop contributions to
$\langle\bar\phi(k_1)\bar\phi(k_2)\bar\phi(k_3)\bar\phi(k_4)\rangle$,
corresponding to diagrams (a)--(f) of Fig.~3 of Ref.~\cite{Frenkel:2026zwc}
(permutations omitted).
}
\label{fig:4pt}
\end{figure}

The one-loop contributions to the four-point function
$\langle\bar\phi\,\bar\phi\,\bar\phi\,\bar\phi\rangle$ are shown in
Fig.~\ref{fig:4pt}.

The figures should be understood as schematic representatives of the
LM assignments associated with the ordinary one-loop topologies of
Ref.~\cite{Frenkel:2026zwc}. The equivalence with the ordinary
one-loop counterterm Lagrangian follows most cleanly from the
functional determinant of the full gauge-fixed field-space operator
$O_{AB}$. Diagrammatically, the same result is represented by mixed
LM propagators, but the admissible diagrams must be constructed from
the full gauge-fixed LM vertex set, rather than by an independent
line-by-line replacement.
Assignments that would require vertices such as
\(\lambda'\sigma'\phi\), \(\lambda'\sigma' h\),
\(\lambda'\lambda'\cdots\), or \(\sigma'\sigma'\cdots\) are excluded.

After the Lee--Yang cancellation the LM functional integral reproduces
the ordinary one-loop counterterm Lagrangian of Ref.~\cite{Frenkel:2026zwc}
(which reduces, for $\xi=\zeta=1$, to the result
obtained by 't~Hooft and Veltman~\cite{tHooft:1974toh}):
\begin{align}
\mathcal{L}_{\rm CT}
&=
\frac{\sqrt{-\bar g}}{16\pi^2\epsilon}
\Big[
\mathcal{C}_1\,\bar R_{\mu\nu}\bar R^{\mu\nu}
+ \mathcal{C}_2\,\bar R^2
+ \kappa^2\mathcal{C}_3\,
\bar R\,\partial_\mu\bar\phi\,\partial^\mu\bar\phi
+ \kappa^2\mathcal{C}_4\,
\bar R_{\mu\nu}\partial^\mu\bar\phi\,\partial^\nu\bar\phi
\nonumber\\
&\quad
+ \kappa^4\mathcal{C}_5\,
\bigl(\partial_\mu\bar\phi\,\partial^\mu\bar\phi\bigr)^2
+ \kappa^2\mathcal{C}_6\,
(\Dbar^2\bar\phi)^2
\Big],
\label{eq:LCT}
\end{align}
where $\epsilon=(4-D)/2$ and the dimensionless gauge-dependent
coefficients $\mathcal{C}_i(\xi,\zeta)$ are given in
Eq.~(3.10) of Ref.~\cite{Frenkel:2026zwc}, rewritten in the
notation used here:
\begin{align}
\mathcal{C}_1
&=
\frac{1}{120}\bigl[43+40\xi(\xi-1)\bigr],
\quad
\mathcal{C}_2
=
\frac{1}{240}\bigl[3+40(\xi-1)^2\bigr],
\label{eq:C1C2}
\\[4pt]
\mathcal{C}_3
&=
-\frac{1}{12}
\bigl[1+(\xi-1)(\xi+5)-3(\xi-\zeta)^2\bigr],
\label{eq:C3}
\\[4pt]
\mathcal{C}_4
&=
\frac{1}{6}(\xi-1)(2\xi+5),
\quad
\mathcal{C}_6
=
\frac{\zeta}{2},
\label{eq:C4C6}
\\[4pt]
\mathcal{C}_5
&=
\frac{1}{8}\bigl[4+2(\xi-1)+(\xi-\zeta)^2\bigr].
\label{eq:C5}
\end{align}
In writing the basis \eqref{eq:LCT} we have used the
four-dimensional Gauss--Bonnet identity to eliminate
$\bar R_{\mu\nu\rho\sigma}\bar R^{\mu\nu\rho\sigma}$, up to a
topological term, and integrations by parts to reduce the scalar
sector to the displayed structures. We use the shorthand
$\Dbar^2\bar\phi \equiv \bar g^{\mu\nu}\Dbar_\mu\Dbar_\nu\bar\phi$,
where $\Dbar_\mu$ is the covariant derivative with respect to
$\bar g_{\mu\nu}$. The structure $(\Dbar^2\bar\phi)^2$ with
coefficient $\mathcal{C}_6$ vanishes on shell by the scalar equation
of motion $\Dbar^2\bar\phi=0$, and is absent in the $\zeta=0$ gauge.
The remaining five structures are the independent curvature and
scalar invariants expected at one loop.

\subsection{The LM field renormalization ansatz}

Since the LM field $\lambda'_{\mu\nu}$ has mass dimension 1 and must
transform as a symmetric tensor under diffeomorphisms, the relevant
local one-loop renormalization for matching the non-redundant
counterterm basis of Eq.~\eqref{eq:LCT} is, up to integrations
by parts and terms associated with scalar equation-of-motion
counterterms,
\begin{equation}
\lambda'_{\mu\nu}
=
\lambda'^r_{\mu\nu}
+
\frac{1}{16\pi^2\epsilon}
\bigl[
\kappa\bigl(c_1\bar R_{\mu\nu}
+ c_2\bar R\,\bar g_{\mu\nu}\bigr)
+
\kappa^3\bigl(c_3\,\partial_\mu\bar\phi\,\partial_\nu\bar\phi
+ c_4\,\bar g_{\mu\nu}\,
\partial_\alpha\bar\phi\,\partial^\alpha\bar\phi\bigr)
\bigr],
\label{eq:lambda-renorm}
\end{equation}
where $c_1, c_2, c_3, c_4$ are dimensionless parameters to be
determined by the matching condition below, and $\lambda'^r_{\mu\nu}$
is the renormalized LM field. This shift gives the bare field
$\lambda'_{\mu\nu}$ a background-dependent component
$\Delta\lambda'_{\mu\nu}[\bar g,\bar\phi]$; the renormalized field
$\lambda'^r_{\mu\nu}$ is the purely quantum degree of freedom.

\subsection{The divergent Lagrangian from LM renormalization}

Substituting Eq.~\eqref{eq:lambda-renorm} into the LM Lagrangian
\eqref{eq:LLM} and using the explicit form of the gravitational
equation-of-motion tensor
$G_{\mu\nu}/\kappa - \kappa T_{\mu\nu}/2$ appearing there,
the shift $\lambda'_{\mu\nu} \to \lambda'^r_{\mu\nu} + \Delta\lambda'_{\mu\nu}$ generates the
following divergent Lagrangian. The gauge-fixing descendant
$\lambda'_{\mu\nu}\,\delta S_{\rm gf}/\delta h_{\mu\nu}$ does not
contribute to the purely background counterterm generated by this
shift, because $\delta S_{\rm gf}/\delta h_{\mu\nu}$ is linear in
the gauge-fixing function $\chi_\mu$ and its derivatives, and
$\chi_\mu$ vanishes when the quantum fields $h_{\mu\nu}$ and
$\varphi$ are set to zero. Thus
the background counterterm matching is governed by the invariant
gravitational equation-of-motion tensor:
\begin{align}
\mathcal{L}_{\rm div}
&=
-\frac{\sqrt{-\bar g}}{16\pi^2\epsilon}
\Bigg[
c_1\,\bar R_{\mu\nu}\bar R^{\mu\nu}
-\Bigl(\frac{c_1}{2}+c_2\Bigr)\bar R^2
\nonumber\\
&\quad
-\kappa^2\!\Bigl(
\frac{c_1}{4}+\frac{c_2}{2}+\frac{c_3}{2}+c_4
\Bigr)
\bar R\,\partial_\mu\bar\phi\,\partial^\mu\bar\phi
\nonumber\\
&\quad
+\kappa^2\!\Bigl(\frac{c_1}{2}+c_3\Bigr)
\bar R_{\mu\nu}\partial^\mu\bar\phi\,\partial^\nu\bar\phi
\nonumber\\
&\quad
+\frac{\kappa^4}{2}\Bigl(\frac{c_3}{2}-c_4\Bigr)
\bigl(\partial_\mu\bar\phi\,\partial^\mu\bar\phi\bigr)^2
\Bigg].
\label{eq:Ldiv}
\end{align}
We note that $\mathcal{L}_{\rm div}$ contains only five structures
and is manifestly free of the $(\Dbar^2\bar\phi)^2$ term.
This is a direct consequence of the fact that the tensor LM shift
$\Delta\lambda'_{\mu\nu}$ multiplies the gravitational equation of
motion. It therefore generates metric equation-of-motion counterterms,
but not scalar equation-of-motion counterterms. The absence of $(\Dbar^2\bar\phi)^2$ from
$\mathcal{L}_{\rm div}$ already signals that the tensor LM
renormalization alone cannot fully absorb $\mathcal{L}_{\rm CT}$.

\subsection{Matching and the residual divergence}

We now match $\mathcal{L}_{\rm div}$ against $\mathcal{L}_{\rm CT}$.
The five structures in Eq.~\eqref{eq:Ldiv} correspond to the first
five structures in Eq.~\eqref{eq:LCT}. Imposing equality on the
four structures $\bar R_{\mu\nu}\bar R^{\mu\nu}$, $\bar R^2$,
$\bar R\,\partial_\mu\bar\phi\,\partial^\mu\bar\phi$, and
$\bar R_{\mu\nu}\partial^\mu\bar\phi\,\partial^\nu\bar\phi$ gives
a linear system whose unique solution is
\begin{align}
c_1 &= -\mathcal{C}_1,
\label{eq:c1}
\\
c_2 &= \frac{\mathcal{C}_1}{2} + \mathcal{C}_2,
\label{eq:c2}
\\
c_3 &= \frac{\mathcal{C}_1}{2} - \mathcal{C}_4,
\label{eq:c3}
\\
c_4 &= -\frac{\mathcal{C}_1}{4}-\frac{\mathcal{C}_2}{2}
       +\mathcal{C}_3+\frac{\mathcal{C}_4}{2}.
\label{eq:c4}
\end{align}
Substituting these values into the coefficient of
$(\partial_\mu\bar\phi\,\partial^\mu\bar\phi)^2$ in
Eq.~\eqref{eq:Ldiv} and defining
\begin{equation}
\Omega
\;\equiv\;
\frac{\mathcal{C}_1}{4}+\frac{\mathcal{C}_2}{4}
-\frac{\mathcal{C}_3}{2}-\frac{\mathcal{C}_4}{2}+\mathcal{C}_5,
\label{eq:Delta-def}
\end{equation}
one finds
\begin{equation}
-\frac{\kappa^4}{2}\Bigl(\frac{c_3}{2}-c_4\Bigr)
=
\kappa^4\Bigl(\mathcal{C}_5 - \Omega\Bigr)
=
\kappa^4\Bigl(\mathcal{C}_5 - \frac{203}{320}\Bigr).
\label{eq:phi4-coeff}
\end{equation}
The explicit verification that $\Omega = 203/320$ for all
$(\xi,\zeta)$ is given in Appendix~\ref{app:matching}.
Therefore, with the optimal choice of $c_i$, the divergent
Lagrangian takes the form
\begin{equation}
\mathcal{L}_{\rm div}
=
\mathcal{L}_{\rm CT}
-
\frac{\sqrt{-\bar g}}{16\pi^2\epsilon}
\left[
\kappa^4
\frac{203}{320}\,
\bigl(\partial_\mu\bar\phi\,\partial^\mu\bar\phi\bigr)^2
+
\kappa^2\mathcal{C}_6\,
(\Dbar^2\bar\phi)^2
\right],
\label{eq:Ldiv-final}
\end{equation}
where the residual coefficient $\Omega = 203/320$ is independent of
both gauge parameters $\xi$ and $\zeta$, as verified explicitly in
Appendix~\ref{app:matching}.

\subsection{Renormalization of the scalar LM field}

The two structures remaining in Eq.~\eqref{eq:Ldiv-final} have
different characters. The first, $\kappa^4(203/320)(\partial\bar\phi)^4$,
is nonzero on shell and is the genuine obstruction to renormalizability,
discussed in Sec.~\ref{sec:obstruction}. The second,
$\kappa^2\mathcal{C}_6(\Dbar^2\bar\phi)^2$, vanishes on shell by
the scalar equation of motion $\Dbar^2\bar\phi=0$.

This second structure is precisely the one generated by renormalizing
the scalar LM field \(\sigma'\). The invariant part of
\(\mathcal{L}_{{\rm LM},\sigma'}\) is
\(\sqrt{-g}\,\sigma' E_\phi\), where
\(E_\phi=-\mathrm{D}^2\phi\). In the purely background sector this
reduces to \(E_{\bar\phi}=-\Dbar^2\bar\phi\).
The most general local one-loop renormalization of \(\sigma'\)
needed for the scalar equation-of-motion counterterm, and consistent
with diffeomorphism invariance and the scalar shift symmetry, is
\begin{equation}
\sigma' = \sigma'^r + \frac{\kappa^2}{16\pi^2\epsilon}\,d\,\Dbar^2\bar\phi,
\label{eq:sigma-renorm}
\end{equation}
where $d$ is a dimensionless parameter and $\sigma'^r$ is the
renormalized scalar LM field. 

The gauge-fixing descendant
\(\sigma'\,\delta S_{\rm gf}/\delta\varphi\) vanishes in the
purely background limit, because it is proportional to \(\chi_\mu\).
Since \(\chi_\mu\) is linear in the quantum fields \(h_{\mu\nu}\)
and \(\varphi\), it vanishes identically when these fields are set
to zero. Therefore this term cannot generate a purely background
counterterm. The background equation-of-motion counterterm generated
by the \(\sigma'\) shift comes only from the invariant part
\(\sqrt{-g}\sigma' E_\phi\).
Substituting into this invariant part of
$S_{{\rm LM},\sigma'}$ gives the leading contribution to the generated
counterterm:
\begin{equation}
\Delta\mathcal{L}_{\sigma'}
=
-
\frac{\sqrt{-\bar g}}{16\pi^2\epsilon}\,
d\,\kappa^2\,(\Dbar^2\bar\phi)^2.
\label{eq:DeltaLsigma}
\end{equation}
The minus sign arises because $E_{\bar\phi} = -\Dbar^2\bar\phi$.
To absorb the residual $+\mathcal{C}_6(\Dbar^2\bar\phi)^2$ in
Eq.~\eqref{eq:Ldiv-final}, we need $-d = \mathcal{C}_6$, i.e.\
\begin{equation}
d = -\mathcal{C}_6 = -\frac{\zeta}{2}.
\label{eq:d-value}
\end{equation}
Note that $d=0$ when $\zeta=0$, consistently with $\mathcal{C}_6=0$
in that gauge. Choosing $d = -\mathcal{C}_6$ absorbs the second
residual in Eq.~\eqref{eq:Ldiv-final} exactly. After both LM
renormalizations, the total divergent Lagrangian reduces to
\begin{equation}
\mathcal{L}_{\rm div} + \Delta\mathcal{L}_{\sigma'}
=
\mathcal{L}_{\rm CT}
-
\frac{\sqrt{-\bar g}\,\kappa^4}{16\pi^2\epsilon}
\frac{203}{320}\,
\bigl(\partial_\mu\bar\phi\,\partial^\mu\bar\phi\bigr)^2.
\label{eq:Ldiv-absorbed}
\end{equation}
The single, gauge-independent obstruction $\kappa^4(203/320)(\partial\bar\phi)^4$
is therefore the only structure that cannot be absorbed by LM field redefinitions.

\paragraph*{On the consistency of the LM field shifts.}
A natural concern is whether substituting the shift
$\sigma' = \sigma'^r + \Delta\sigma'$ into the \emph{full} quantum action
--- not only into the purely background part --- generates new interaction
vertices, and hence possibly new divergent contributions, in the
reorganized perturbative expansion.

To appreciate why such a reorganization is harmless, consider the
following elementary example. Suppose a divergent quantity
$T_{\rm div}$ appears in a combination $T_{\rm div} + \lambda'$, where
$\lambda'$ is itself divergent. Defining a renormalized field
$\lambda' = \lambda'^r - T_{\rm div}$ rewrites the combination as
$\lambda'^r$, which is finite: two individually divergent contributions
have been reorganized into a single finite expression without any new
divergence being introduced. The role of the LM field shift below is
precisely analogous.

The shift $\Delta\sigma'[\bar g,\bar\phi]$ depends only on the background
fields and is independent of the quantum integration variable $\sigma'$.
It is therefore a constant translation in the path-integral measure,
$\mathcal{D}\sigma' = \mathcal{D}\sigma'^r$, with trivial unit Jacobian.
This shows that the change of variables is admissible in the exact
functional integral; by itself, however, it does not establish that the
vertices induced by the shift give no net contribution at one loop, since
the same translation reorganizes the action and redistributes
contributions among the diagrams.

Alternatively, these induced vertices may be viewed as counterterm
insertions generated by an \(O(\hbar)\) field redefinition. The more
direct diagrammatic reason why they do not generate new loop UV
divergences is the absence of vertices bilinear in the LM fields, as
explained below and illustrated in Appendix~\ref{app:phi3}.

That this mechanism indeed operates is confirmed explicitly in the scalar
toy model of Appendix~\ref{app:phi3}.
The same structural counting applies in the LM formulation of quantum
gravity and is not restricted to one loop.
Since there are no terms quadratic in the LM fields, the vertices
generated by the background-dependent parts of the LM field shifts
contain background-field factors multiplying ordinary non-LM quantum
fields, such as $\varphi$ and $h_{\mu\nu}$, but they do not generate
vertices with two LM legs.
Every internal line attached to such a vertex must therefore be a
mixed propagator, with its LM end absorbed at a subsequent vertex.
Since no vertex in the LM theory carries two LM legs
(Eq.~\eqref{eq:absent-vertices}), no such chain of propagators can
ever close into a loop.
The induced vertices therefore do not generate new loop UV divergences;
their only role is to reorganize counterterm insertions already
introduced by the LM field renormalizations, and they introduce no new
divergences at any order of perturbation theory.

The Faddeev--Popov ghost action is not affected by these LM field
shifts, since it contains no direct coupling to $\lambda'_{\mu\nu}$ or
$\sigma'$. Therefore the LM renormalizations generate no additional
ghost vertices.

This result corresponds to the on-shell counterterm
\begin{equation}
\mathcal{L}_{\rm CT}\big|_{\rm on\text{-}shell}
=
\frac{\sqrt{-\bar g}\,\kappa^4}{16\pi^2\epsilon}
\frac{203}{320}
\bigl(\partial_\mu\bar\phi\,\partial^\mu\bar\phi\bigr)^2,
\label{eq:onshell-CT}
\end{equation}
obtained by substituting the on-shell relations
\begin{equation}
\bar R_{\mu\nu}
=
-\frac{\kappa^2}{2}
\partial_\mu\bar\phi\,\partial_\nu\bar\phi,
\quad
\bar R
=
-\frac{\kappa^2}{2}
\partial_\mu\bar\phi\,\partial^\mu\bar\phi,
\quad
\Dbar^2\bar\phi=0,
\label{eq:on-shell-relations}
\end{equation}
into Eq.~\eqref{eq:LCT} and using the definition
\eqref{eq:Delta-def} of \(\Omega=203/320\), verified in
Appendix~\ref{app:matching}. The signs in
Eq.~\eqref{eq:on-shell-relations} follow from the convention for
\(T_{\mu\nu}\) in Eq.~\eqref{eq:Tmunu}.

\section{The obstruction to renormalizability}
\label{sec:obstruction}

\subsection{Physical character of the residual obstruction}

The result \eqref{eq:Ldiv-absorbed} shows that, after renormalization
of both LM fields $\lambda'_{\mu\nu}$ and $\sigma'$, a single structure
remains unmatched: the gauge-independent term
$\kappa^4(203/320)(\partial\bar\phi)^4$.

The on-shell-vanishing structure $\kappa^2\mathcal{C}_6(\Dbar^2\bar\phi)^2$
is completely absorbed by the scalar LM renormalization
\eqref{eq:sigma-renorm}, as shown in Sec.~\ref{sec:CT}. It leaves no
residual on shell and does not contribute to physical amplitudes.

The remaining residual, $\kappa^4(203/320)(\partial\bar\phi)^4$, is the
genuine obstruction. It is precisely the on-shell value of
$\mathcal{L}_{\rm CT}$, Eq.~\eqref{eq:onshell-CT}, which is
nonzero and gauge-independent. The combined LM renormalization
therefore fails to absorb exactly the physical on-shell divergence of
the theory.

\subsection{Structural argument: equation-of-motion counterterms}

The residual $\Omega = 203/320$ is not a limitation of the
particular ansatz \eqref{eq:lambda-renorm}. It is a structural
consequence of the way $\lambda'_{\mu\nu}$ enters the action.

Consider the most general local shift of the LM field,
\begin{equation}
\lambda'_{\mu\nu}
\;\longrightarrow\;
\lambda'^r_{\mu\nu}
+
\Delta\lambda'_{\mu\nu}[\bar g,\bar\phi],
\label{eq:general-shift}
\end{equation}
where $\Delta\lambda'_{\mu\nu}[\bar g,\bar\phi]$ is any local
functional of the background fields built from curvature tensors,
scalar gradients, and their covariant derivatives. Substituting
into the LM Lagrangian \eqref{eq:LLM}, the change in the action is
\begin{equation}
\Delta S_{\rm LM}
=
-\int d^4x\,\sqrt{-\bar g}\,
\Delta\lambda'^{\mu\nu}[\bar g,\bar\phi]\,
\left[
\frac{1}{\kappa}
\!\left(
\bar R_{\mu\nu}-\frac{1}{2}\bar R\,\bar g_{\mu\nu}
\right)
-
\frac{\kappa}{2}\bar T_{\mu\nu}
\right].
\label{eq:DeltaSLM}
\end{equation}
Every term in Eq.~\eqref{eq:DeltaSLM} is proportional to the
classical gravitational equation of motion
$\bar R_{\mu\nu} - \frac{1}{2}\bar R\,\bar g_{\mu\nu}
- \frac{\kappa^2}{2}\bar T_{\mu\nu}= 0$.
Although the LM fields are coupled to the gauge-fixed quantum
equations in the perturbative formulation, the gauge-fixing
descendants vanish on physical backgrounds with the quantum fields
set to zero. Therefore the on-shell background counterterms generated
by LM renormalizations are still equation-of-motion counterterms of
the invariant classical theory. Such terms vanish identically when
the background fields satisfy the equations of motion, i.e.\ on shell.

This is a direct structural consequence of the fact that
$\lambda'_{\mu\nu}$ enters the LM Lagrangian \eqref{eq:LLM}
\emph{linearly}, multiplied by the gravitational equation of motion.
No matter how complicated
$\Delta\lambda'_{\mu\nu}[\bar g,\bar\phi]$ is, the resulting
counterterm \eqref{eq:DeltaSLM} will always vanish on the
gravitational mass shell.

Similarly, the most general local shift of the scalar LM field,
\begin{equation}
\sigma'
\;\longrightarrow\;
\sigma'^r + \Delta\sigma'[\bar g,\bar\phi],
\label{eq:general-sigma-shift}
\end{equation}
substituted into the invariant scalar LM coupling
$\sqrt{-g}\,\sigma' E_\phi$, generates
\begin{equation}
\Delta S_{\sigma'}
=
\int d^4x\,\sqrt{-\bar g}\,
\Delta\sigma'[\bar g,\bar\phi]\,E_{\bar\phi},
\label{eq:DeltaSsigma}
\end{equation}
which vanishes identically on the scalar mass shell $E_{\bar\phi}
= -\Dbar^2\bar\phi = 0$.

Thus $\lambda'_{\mu\nu}$ shifts generate gravitational
equation-of-motion counterterms, while $\sigma'$ shifts generate
scalar equation-of-motion counterterms. Neither class can cancel a
genuine on-shell invariant such as $\kappa^4(\partial\bar\phi)^4$,
which is nonzero when both the gravitational and scalar equations of
motion are satisfied.

Since equation-of-motion counterterms vanish on shell, and since
the on-shell counterterm \eqref{eq:onshell-CT} is nonzero, it
follows that a local renormalization of the LM field
$\lambda'_{\mu\nu}$ cannot remove the on-shell divergence
\eqref{eq:onshell-CT}. The obstruction $\Omega = 203/320$ is a
genuine gauge-independent on-shell obstruction, independent of any
choice of $\Delta\lambda'_{\mu\nu}[\bar g,\bar\phi]$.

\subsection{Gauge-independence of the obstruction}

The gauge-independence of the residual coefficient $203/320$
is consistent with the DeWitt--Kallosh theorem~\cite{DeWitt:1967yk,Kallosh:1974yh}. The on-shell
combination
\begin{equation}
\frac{\mathcal{C}_1}{4}+\frac{\mathcal{C}_2}{4}
-\frac{\mathcal{C}_3}{2}-\frac{\mathcal{C}_4}{2}+\mathcal{C}_5
= \frac{203}{320}
\label{eq:Delta-combo}
\end{equation}
is verified explicitly in Appendix~\ref{app:matching} to be
independent of both $\xi$ and $\zeta$:
\begin{equation}
\frac{\partial}{\partial\xi}
\Bigl(
\frac{\mathcal{C}_1}{4}+\frac{\mathcal{C}_2}{4}
-\frac{\mathcal{C}_3}{2}-\frac{\mathcal{C}_4}{2}+\mathcal{C}_5
\Bigr)
= 0,
\qquad
\frac{\partial}{\partial\zeta}
\Bigl(
\frac{\mathcal{C}_1}{4}+\frac{\mathcal{C}_2}{4}
-\frac{\mathcal{C}_3}{2}-\frac{\mathcal{C}_4}{2}+\mathcal{C}_5
\Bigr)
= 0.
\label{eq:Delta-gauge-indep}
\end{equation}
The two-parameter gauge analysis shows that the obstruction is
independent of the gauge parameters $\xi$ and $\zeta$. It
therefore does not rely on a special choice of background gauge.
There is no choice of $(\xi,\zeta)$ that makes the residual vanish.

\subsection{Connection to the renormalizability condition}

The algebraic content of this obstruction is captured by the
consistency condition for renormalizability by field redefinitions
derived in Ref.~\cite{Frenkel:2026zwc}: all divergences can be
absorbed by redefinitions of $g_{\mu\nu}$ and $\phi$ if and only
if the on-shell counterterm vanishes, which requires
\begin{equation}
\frac{\mathcal{C}_1}{4}+\frac{\mathcal{C}_2}{4}
-\frac{\mathcal{C}_3}{2}-\frac{\mathcal{C}_4}{2}+\mathcal{C}_5
= 0.
\label{eq:consistency}
\end{equation}
Since the left-hand side equals $203/320 \neq 0$ for all
$(\xi,\zeta)$, the theory is not renormalizable by field
redefinitions, and the LM renormalization cannot change this
conclusion: a shift $\Delta\lambda'_{\mu\nu}$ produces only
equation-of-motion counterterms, which make no contribution to
the on-shell effective action.

\section{Pure gravity versus gravity coupled to scalar matter}
\label{sec:puregravity}

\subsection{Pure quantum gravity: LM renormalization alone suffices}

The result of the previous section shows that in gravity coupled to scalar matter 
the LM renormalization leaves a genuine on-shell residual
$\kappa^4(203/320)(\partial\bar\phi)^4$. It is natural to ask
whether this obstruction is a consequence of the matter sector, or
whether it would persist even in the absence of the scalar field.

In pure quantum gravity (no matter, $\phi = 0$), the one-loop
counterterm Lagrangian in a given background gauge takes the
form~\cite{tHooft:1974toh,Grisaru:1975ei}
\begin{equation}
\mathcal{L}_{\rm CT}^{\rm grav}
=
\frac{\sqrt{-\bar g}}{16\pi^2\epsilon}
\left[
A_1(\xi)\,\bar R_{\mu\nu}\bar R^{\mu\nu}
+A_2(\xi)\,\bar R^2
\right],
\label{eq:LCT-grav}
\end{equation}
where $A_1(\xi)$ and $A_2(\xi)$ are gauge-dependent coefficients.
In four dimensions, the third possible curvature-squared structure,
$\bar R_{\mu\nu\rho\sigma}\bar R^{\mu\nu\rho\sigma}$, may be
traded for $\bar R_{\mu\nu}\bar R^{\mu\nu}$ and $\bar R^2$ using
the Gauss--Bonnet identity
$\bar R_{\mu\nu\rho\sigma}\bar R^{\mu\nu\rho\sigma}
-4\bar R_{\mu\nu}\bar R^{\mu\nu}+\bar R^2 = \text{total derivative}$,
so the two-structure basis \eqref{eq:LCT-grav} is complete.
The matter-dependent structures are absent, and in particular
there is no $(\partial\bar\phi)^4$ term.

The LM renormalization ansatz in the gravitational sector reduces
to
\begin{equation}
\lambda'_{\mu\nu}
=
\lambda'^r_{\mu\nu}
+
\frac{\kappa}{16\pi^2\epsilon}
\bigl(c_1\bar R_{\mu\nu} + c_2\bar R\,\bar g_{\mu\nu}\bigr),
\label{eq:lambda-renorm-grav}
\end{equation}
which generates the divergent Lagrangian
\begin{equation}
\mathcal{L}_{\rm div}^{\rm grav}
=
-\frac{\sqrt{-\bar g}}{16\pi^2\epsilon}
\left[
c_1\,\bar R_{\mu\nu}\bar R^{\mu\nu}
-\Bigl(\frac{c_1}{2}+c_2\Bigr)\bar R^2
\right].
\label{eq:Ldiv-grav}
\end{equation}
Matching against Eq.~\eqref{eq:LCT-grav} gives
\begin{equation}
c_1 = -A_1(\xi),
\qquad
c_2 = \frac{A_1(\xi)}{2} + A_2(\xi),
\label{eq:c1c2-grav}
\end{equation}
and with this choice $\mathcal{L}_{\rm div}^{\rm grav} =
\mathcal{L}_{\rm CT}^{\rm grav}$ exactly. There is no residual.

We therefore conclude that in pure quantum gravity at one loop,
the LM field renormalization \eqref{eq:lambda-renorm-grav} with
the choice \eqref{eq:c1c2-grav} absorbs all one-loop divergences
and no additional non-topological counterterms are required. This
has a simple physical interpretation: up to the Gauss--Bonnet
topological density, the one-loop divergences of pure gravity are
curvature-squared terms generated by local field redefinitions
proportional to the vacuum Einstein equations
$\bar R_{\mu\nu} - \frac{1}{2}\bar R\,\bar g_{\mu\nu} = 0$ --- precisely
the equation-of-motion counterterms that the LM renormalization
generates, allowing complete cancellation. For the one-loop model of
gravity coupled to scalar matter considered here, the coupling to a
scalar field instead makes the on-shell counterterm nonzero,
$\kappa^4(203/320)(\partial\bar\phi)^4$, and no equation-of-motion
counterterm can cancel it.

\subsection{Quantum gravity with scalar matter: one additional counterterm required}

In the gravity coupled to scalar matter theory, the LM renormalization
of $\lambda'_{\mu\nu}$ and $\sigma'$ absorbs all one-loop divergences
except the single irreducible residual
$\kappa^4(203/320)(\partial\bar\phi)^4$. To
cancel this remaining one-loop divergence, one additional counterterm must
be introduced:
\begin{equation}
\mathcal{L}_{\rm add}
=
\frac{\sqrt{-\bar g}\,\kappa^4}{16\pi^2\epsilon}
\frac{203}{320}\,
\bigl(\partial_\mu\bar\phi\,\partial^\mu\bar\phi\bigr)^2.
\label{eq:Ladd}
\end{equation}
Together with the LM renormalization of $\lambda'_{\mu\nu}$ and
$\sigma'$, this counterterm cancels all one-loop divergences of the
matter-coupled theory considered here.

This additional structure has the correct mass dimension: the operator
$(\partial\bar\phi)^4$ has mass dimension 8 and the accompanying
factor $\kappa^4$ has dimension $-4$, so the Lagrangian-density
term $\kappa^4(\partial\bar\phi)^4$ has dimension 4, as required.

The theory is therefore renormalized at one loop in the effective
field theory sense: all one-loop divergences of the matter-coupled
theory can be cancelled by the LM field renormalizations together
with the single matter counterterm \eqref{eq:Ladd}.

Within the one-loop analysis, the LM field redefinitions remove all
counterterms proportional to the background equations of motion, while
the remaining gauge-independent obstruction is cancelled only after
adding the independent operator \eqref{eq:Ladd}. The one-loop exactness
of the LM formulation, established in
Refs.~\cite{Brandt:2023lyq,McKeon:2024psy} and discussed in
Sec.~\ref{sec:formulation}, suggests that, once the extended action is
treated consistently, no ordinary tower of higher-loop radiative
corrections is expected. This statement should be understood with the
qualification that the LM fields must be coupled to the extended action
and that the vertices induced by the LM field shifts must be controlled,
as discussed in Sec.~\ref{sec:CT} and Appendix~\ref{app:phi3}. In
particular, both LM fields are genuinely required once matter is
present: a single tensor multiplier $\lambda'_{\mu\nu}$, coupled to
matter but without the scalar multiplier $\sigma'$, already fails to
enforce one-loop exactness in the mixed gravity--matter sector beyond
one loop, as illustrated explicitly at two-loop order in
Appendix~\ref{app:twoloop}. This is at variance with the claim made in
Ref.~\cite{Chishtie:2025}.

\section{Conclusion}
\label{sec:conclusion}

We have studied the one-loop renormalization of Einstein gravity
coupled to a scalar field in the gauge-fixed LM
formulation. The formulation introduces a pair of quantum
Lagrange multiplier fields, $(\lambda'_{\mu\nu},\sigma')$, coupled to
the gauge-fixed gravitational and scalar equations of motion
respectively. In the background-field method with the general
two-parameter gauge of Ref.~\cite{Frenkel:2026zwc}, the quadratic
LM matrix takes the canonical block form with the full gauge-fixed
Hessian $O_{AB}$ in both the ordinary and mixed blocks. The
determinant doubling is cancelled by Lee--Yang ghost fields acting
on $O_{AB}$, establishing the one-loop equivalence of the LM and
ordinary theories. The main results are as follows.

The tensor LM renormalization, Eq.~\eqref{eq:lambda-renorm},
uniquely fixes four coefficients $c_1,\ldots,c_4$ by absorbing the
gravitational and mixed counterterm structures
$\bar R_{\mu\nu}\bar R^{\mu\nu}$, $\bar R^2$,
$\bar R\,\partial_\mu\bar\phi\,\partial^\mu\bar\phi$, and
$\bar R_{\mu\nu}\partial^\mu\bar\phi\,\partial^\nu\bar\phi$.
The scalar LM renormalization, Eq.~\eqref{eq:sigma-renorm} with
$d = -\mathcal{C}_6 = -\zeta/2$, subsequently absorbs the
equation-of-motion structure $\kappa^2(\zeta/2)(\Dbar^2\bar\phi)^2$.
This structure is present only when $\zeta\neq0$ and vanishes on
shell by the scalar equation of motion $\Dbar^2\bar\phi=0$; it is
a gauge artefact rather than a physical divergence.

The consistency of both LM renormalizations---the absence of new physical
divergences induced by the field shifts at the quantum level---follows,
under the one-loop exactness of the LM formulation, from a loop-counting
argument confirmed explicitly in the scalar toy model of
Appendix~\ref{app:phi3}, as discussed in Sec.~\ref{sec:CT}.

After both LM renormalizations, all off-shell counterterms proportional
to the background equations of motion are absorbed into the LM field
redefinitions. The only remaining off-shell counterterm, proportional
to $(\partial_\mu\bar\phi\,\partial^\mu\bar\phi)^2$, does not vanish
on shell and therefore cannot be absorbed by any further LM
renormalization: it is the single independent shift-symmetric operator
left over. Therefore the LM formulation does not make the original
scalar-gravity action finite by itself; rather,
it reduces the required additional counterterm to this single residual
structure. This is
in contrast with pure gravity, where all one-loop counterterms are
proportional to the vacuum Einstein equations of motion; in this case,
the renormalization of $\lambda^\prime_{\mu\nu}$ absorbs all divergences
which are consistent with the symmetry of the theory.

The scalar LM field $\sigma'$, required to enforce one-loop exactness
in the presence of both gravitational and matter fields
(Sec.~\ref{sec:puregravity},
Appendix~\ref{app:twoloop}), modifies the matter propagator structure,
which appears to obstruct a direct unification of the
present formulation of gravity with the perturbative Standard Model.
This limitation reflects the limited domain of validity of the present
low-energy theories of gravity and of the Standard Model. A genuine
unification of gravity and the Standard Model should occur only at
energies far above those currently accessible, when both theories in
their present low-energy forms will be substantially modified.

It would be worthwhile to extend the present treatment to models
including more general couplings between quantum gravity and matter
fields~\cite{Frenkel:2026zwc}. Another interesting generalization
involves the analysis of the Einstein--Cartan theory, which possesses
two distinct local gauge invariances---diffeomorphisms and local
Lorentz
transformations~\cite{Hehl:2024gauge,Brandt:2024aop1,Brandt:2024aop2}.
These issues are currently being examined.

\begin{acknowledgments}
D.\ G.\ C\@. M\@. thanks Ann Aksoy for enlightening conversations.
F.\ T.\ B\@. and J.\ F\@. thank CNPq (Brazil) for financial support.
S.\ M.-F\@. thanks FAPESP for financial support. This study was
financed, in part, by the S\~ao Paulo Research Foundation (FAPESP),
Brasil. Process Number \#2025/16156-7.
\end{acknowledgments}
\appendix
\renewcommand{\theequation}{\thesection.\arabic{equation}}

\section{Feynman rules}
\label{app:diagrams}

This appendix collects the propagators and vertices needed to
evaluate the one-loop diagrams corresponding to Figs.~1--3 of
Ref.~\cite{Frenkel:2026zwc}. Each rule is derived from the
Lagrangians of Sec.~\ref{sec:formulation}; the relevant action
terms are indicated for each entry. The diagrammatic application
of these rules---the one-loop graphs with mixed internal lines and
their relation to the results of Ref.~\cite{Frenkel:2026zwc}---is
discussed in Sec.~\ref{sec:CT}.
We employ the following graphic conventions:
\begin{center}
\renewcommand{\arraystretch}{2.2}
\begin{tabular}{ll}
\propfig{bgrav}{$\bar h_{\mu\nu}$}\\[1mm]
\begin{tikzpicture}[baseline=-0.5ex,line cap=round,line join=round]
  \mixedWD{0}{0}{2.2}{0}
\end{tikzpicture}
\;=\;
$\displaystyle\langle h_{\mu\nu}\lambda'_{\rho\sigma}\rangle_q
= \langle\lambda'_{\mu\nu}h_{\rho\sigma}\rangle_q
= -\frac{iP_{\mu\nu,\rho\sigma}}{k^2+i\varepsilon}
= -\langle\lambda'_{\mu\nu}\lambda'_{\rho\sigma}\rangle_q$
\\[3mm]
\begin{tikzpicture}[baseline=-0.5ex,line cap=round,line join=round]
  \draw[ghost,postaction={decorate,decoration={markings,
    mark=at position 0.5 with{\arrow[>=stealth]{>}}}}]
    (0,0)--(2.2,0);
\end{tikzpicture}
\;=\;
$\displaystyle\langle c^\mu\bar c_\nu\rangle_q
= \frac{i\delta^\mu{}_\nu}{k^2+i\varepsilon}$
\\[3mm]
\propfig{bscalar}{$\bar\phi$}\\[1mm]
\begin{tikzpicture}[baseline=-0.5ex,line cap=round,line join=round]
  \mixedSscalar{0}{0}{2.2}{0}
\end{tikzpicture}
\;=\;
$\displaystyle\langle\varphi\,\sigma'\rangle_q
= \langle\sigma'\,\varphi\rangle_q
= +\frac{i}{k^2+i\varepsilon}
= -\langle\sigma'\,\sigma'\rangle_q$
\end{tabular}
\end{center}

\subsection{Propagators}
\label{app:props}

We work around a flat background ($\bar g_{\mu\nu}=\eta_{\mu\nu}$).
Since $\bar\lambda_{\mu\nu}=\bar\sigma=0$
(Eqs.~\eqref{eq:lambda-purely-quantum} and
\eqref{eq:sigma-purely-quantum}), there are no background LM lines.
The nonvanishing propagators of the quantum fields follow from the
quadratic parts of the actions
$S_g+S_{\rm gf}$ (Eqs.~\eqref{eq:Sg}, \eqref{eq:Lgf}),
$S_{\rm LM}$ (Eq.~\eqref{eq:SLM}),
$S_\phi+S_{{\rm LM},\sigma'}$ (Eqs.~\eqref{eq:Sphi},
\eqref{eq:SLMsigma}), and $S_{\rm gh}$
(Eq.~\eqref{eq:Lgh}).

\noindent
The vanishing of $\langle hh\rangle_q$ and $\langle\varphi\varphi\rangle_q$
follows from the block structure of the LM quadratic forms
(Sec.~\ref{sec:formulation}). The full chain of equalities for the
gravitational sector,
\begin{equation}
\langle\lambda'_{\mu\nu}\lambda'_{\rho\sigma}\rangle_q
= -\langle h_{\mu\nu}\lambda'_{\rho\sigma}\rangle_q
= -\langle\lambda'_{\mu\nu}h_{\rho\sigma}\rangle_q
= \frac{i\,P_{\mu\nu,\rho\sigma}(k,\xi,D)}{k^2+i\varepsilon},
\label{eq:app-prop-rel}
\end{equation}
and for the scalar sector,
\begin{equation}
\langle\sigma'\,\sigma'\rangle_q
= -\langle\varphi\,\sigma'\rangle_q
= -\langle\sigma'\,\varphi\rangle_q
= -\frac{i}{k^2+i\varepsilon},
\label{eq:app-prop-rel-scalar}
\end{equation}
follow from Eqs.~\eqref{eq:propagator-relation} and
\eqref{eq:propagator-relation-scalar} of Sec.~\ref{sec:formulation}.
Here $P_{\mu\nu,\rho\sigma}(k,\xi,D)$ is the gauge-dependent spin-2
projector given in Eq.~(A12) of Ref.~\cite{Frenkel:2026zwc}. In the
de~Donder gauge ($\xi=1$) at $D=4$ it reduces to
\begin{equation}
P_{\mu\nu,\rho\sigma}
=
\tfrac{1}{2}
\!\left(
\eta_{\mu\rho}\eta_{\nu\sigma}
+\eta_{\mu\sigma}\eta_{\nu\rho}
-\eta_{\mu\nu}\eta_{\rho\sigma}
\right).
\label{eq:P-deDonder}
\end{equation}

\subsection{Vertices}
\label{app:verts}

All vertices are derived from the gauge-fixed LM action
\eqref{eq:SLMtot} of Sec.~\ref{sec:formulation}. For the invariant
scalar kinetic part, the key identities
$V_{\lambda'\phi\phi}=V_{h\phi\phi}=V_{\bar h\phi\phi}$
(Sec.~\ref{sec:formulation}) and the matter-multiplier analogue
$V_{H\sigma'\phi}=V_{H\phi\phi}$ (Eq.~\eqref{eq:sigma-vertex})
reduce all graviton--scalar--scalar vertices to the same $\tau^{\mu\nu}$ form.
In the full gauge-fixed theory these identities are supplemented by
the gauge-fixing descendant vertices generated by
$\lambda'_{\mu\nu}\,\delta S_{\rm gf}/\delta h_{\mu\nu}$ and
$\sigma'\,\delta S_{\rm gf}/\delta\varphi$, which are listed separately below.
The $\tau^{\mu\nu}$ vertices from the invariant kinetic term are
(Eqs.~\eqref{eq:lambda-phi-phi} and \eqref{eq:sigma-vertex} of
Sec.~\ref{sec:formulation}):
\begin{align}
&\threevertex{bscalar}{bscalar}{bgrav}{\bar\phi}{\bar\phi}{\bar h_{\mu\nu}}
\;=\;
\threevertex{bscalar}{qscalar}{bgrav}{\bar\phi}{\varphi}{\bar h_{\mu\nu}}
\;=\;
\threevertex{qscalar}{qscalar}{bgrav}{\varphi}{\varphi}{\bar h_{\mu\nu}}
\;=\; -\frac{i\kappa}{2}\tau^{\mu\nu}(p,q),
\nonumber\\[4mm]
&\threevertex{bscalar}{bscalar}{qgrav}{\bar\phi}{\bar\phi}{h_{\mu\nu}}
\;=\;
\threevertex{bscalar}{qscalar}{qgrav}{\bar\phi}{\varphi}{h_{\mu\nu}}
\;=\;
\threevertex{qscalar}{qscalar}{qgrav}{\varphi}{\varphi}{h_{\mu\nu}}
\;=\; -\frac{i\kappa}{2}\tau^{\mu\nu}(p,q),
\nonumber\\[4mm]
&\threevertex{bscalar}{bscalar}{qlm}{\bar\phi}{\bar\phi}{\lambda'_{\mu\nu}}
\;=\;
\threevertex{bscalar}{qscalar}{qlm}{\bar\phi}{\varphi}{\lambda'_{\mu\nu}}
\;=\;
\threevertex{qscalar}{qscalar}{qlm}{\varphi}{\varphi}{\lambda'_{\mu\nu}}
\;=\; -\frac{i\kappa}{2}\tau^{\mu\nu}(p,q),
\nonumber\\[4mm]
&\threevertex{qslm}{bscalar}{bgrav}{\sigma'}{\bar\phi}{\bar h_{\mu\nu}}
\;=\;
\threevertex{qslm}{qscalar}{bgrav}{\sigma'}{\varphi}{\bar h_{\mu\nu}}
\;=\;
\threevertex{qslm}{bscalar}{qgrav}{\sigma'}{\bar\phi}{h_{\mu\nu}}
\nonumber\\
&\;=\;
\threevertex{qslm}{qscalar}{qgrav}{\sigma'}{\varphi}{h_{\mu\nu}}
\;=\; -\frac{i\kappa}{2}\tau^{\mu\nu}(p,q),
\label{eq:cubic-scalar-vertices}
\end{align}
where $\tau^{\mu\nu}(p,q)=p^\mu q^\nu+p^\nu q^\mu-\eta^{\mu\nu}p\cdot q$
(massless case) and $p$, $q$ are the two scalar momenta. These cubic
vertices arise from $S_\phi$ (Eq.~\eqref{eq:Sphi}), $S_{\rm LM}$
(Eq.~\eqref{eq:SLM}), and $S_{{\rm LM},\sigma'}$
(Eq.~\eqref{eq:SLMsigma}).

The LM gravitational cubic vertex is obtained from the ordinary
Einstein--Hilbert three-graviton vertex by designating one leg as
$\lambda'_{\mu\nu}$, derived from $S_{\rm LM}$ (Eq.~\eqref{eq:SLM}):
\begin{align}
&\threevertex{qlm}{bgrav}{bgrav}{\lambda'}{\bar h}{\bar h}
\;=\;
\threevertex{qlm}{bgrav}{qgrav}{\lambda'}{\bar h}{h}
\;=\;
\threevertex{qlm}{qgrav}{qgrav}{\lambda'}{h}{h}
\;=\; V^{(3)\,\mu\nu,\rho\sigma,\alpha\beta}_{\rm EH}(k_1,k_2,k_3)
\Big|_{\bar h_{\mu\nu}\to\lambda'_{\mu\nu}},
\label{eq:cubic-grav-lm}
\end{align}
where
\(\big|_{\bar h_{\mu\nu}\to\lambda'_{\mu\nu}}\)
means that the graviton leg carrying the indices \(\mu\nu\) is
replaced by a \(\lambda'_{\mu\nu}\) leg, without symmetrization over
that leg (Eq.~\eqref{eq:lambda-grav-rule}).

For the quartic vertices, the two-graviton--two-scalar seagull
and its LM/$\sigma'$ descendants (from $S_\phi$, $S_{\rm LM}$,
$S_{{\rm LM},\sigma'}$):
\begin{align}
&\fourvertex{bgrav}{qgrav}{bscalar}{qscalar}
  {\bar h_{\mu\nu}}{h_{\rho\sigma}}{\bar\phi}{\varphi}
\;=\;
\fourvertex{bgrav}{qlm}{bscalar}{qscalar}
  {\bar h_{\mu\nu}}{\lambda'_{\rho\sigma}}{\bar\phi}{\varphi}
\;=\;
\fourvertex{bgrav}{qgrav}{bscalar}{qslm}
  {\bar h_{\mu\nu}}{h_{\rho\sigma}}{\bar\phi}{\sigma'}
=\;
\frac{i\kappa^2}{2}\,\Theta^{\mu\nu,\rho\sigma}(p,q),
\label{eq:seagull-vertex}
\end{align}
where $\Theta^{\mu\nu,\rho\sigma}(p,q)$ is the contact structure
obtained by expanding $\sqrt{-g}\,g^{\alpha\beta}\partial_\alpha\phi\,
\partial_\beta\phi$ to second order in the graviton field; it is
symmetric under $\mu\!\leftrightarrow\!\nu$,
$\rho\!\leftrightarrow\!\sigma$,
$(\mu\nu)\!\leftrightarrow\!(\rho\sigma)$, and
$p\!\leftrightarrow\!q$; the explicit form is given in Eq.~(A27)
of Ref.~\cite{Frenkel:2026zwc}.

The four-graviton vertex and its single-$\lambda'$ descendant (from
$S_g$ and $S_{\rm LM}$, Eqs.~\eqref{eq:Sg} and
\eqref{eq:SLM}):
\begin{align}
&\fourvertex{bgrav}{qgrav}{bgrav}{qgrav}
  {\bar h_{\mu\nu}}{h_{\rho\sigma}}{\bar h_{\alpha\beta}}{h_{\gamma\delta}}
=
V^{(4)\,\mu\nu,\rho\sigma,\alpha\beta,\gamma\delta}_{\rm EH},
\nonumber\\
&\fourvertex{qlm}{qgrav}{bgrav}{qgrav}
  {\lambda'_{\mu\nu}}{h_{\rho\sigma}}{\bar h_{\alpha\beta}}{h_{\gamma\delta}}
=
\left.
V^{(4)\,\mu\nu,\rho\sigma,\alpha\beta,\gamma\delta}_{\rm EH}
\right|_{\bar h_{\mu\nu}\to\lambda'_{\mu\nu}} .
\label{eq:fourgraviton-vertex}
\end{align}
where the $\lambda'$ leg is not symmetrized with the graviton legs.

The ghost--graviton vertex $V_{\bar c\,\bar h\,c}$ (from $S_{\rm gh}$,
Eq.~\eqref{eq:Lgh}) is the standard one~\cite{Frenkel:2026zwc}.

The gauge-fixing descendant of $\sigma'$
(from $\sigma'\,\delta S_{\rm gf}/\delta\varphi$,
Eq.~\eqref{eq:LLMsigma}) generates two additional vertex classes
proportional to $\zeta$. The formulas below display the momentum and
index structures in the conventions of Eq.~\eqref{eq:LLMsigma}; the
overall factor of $i$ is to be included when converting them into
Feynman-rule vertices. Expanding the gauge-fixing function
$\chi^\mu = h^{\mu\nu}{}_{;\nu}-\tfrac{1}{2}h^\alpha{}_{\alpha;}{}^\mu
-\zeta\kappa\varphi\,\partial^\mu\bar\phi$, the new vertices are:
\begin{align}
V_{\sigma' h\partial\bar\phi}
&= \frac{\zeta\kappa}{\xi}
   \Bigl(
     k_1^{\mu}\delta^\nu{}_\rho
     +k_1^{\nu}\delta^\mu{}_\rho
     -\eta^{\mu\nu}k_{1\,\rho}
   \Bigr)\partial^\rho\bar\phi
   \quad(\sigma',\,h_{\mu\nu},\,\bar\phi\text{ vertex}),
\label{eq:sigma-gf-vertex1}
\\[4pt]
V_{\sigma'\varphi(\partial\bar\phi)^2}
&= -\frac{\zeta^2\kappa^2}{\xi}
   \,(\partial\bar\phi)^2
   \quad(\sigma',\,\varphi,\,\bar\phi^2\text{ vertex}).
\label{eq:sigma-gf-vertex2}
\end{align}
Both vanish when $\zeta=0$, consistently with $\mathcal{C}_6=\zeta/2\to 0$
in that gauge.

Because $S_{\rm LM}^{\rm tot}$ is linear in $\Lambda_A=(\lambda'_{\mu\nu},\sigma')$,
every vertex carries at most one
$\lambda'_{\mu\nu}$ leg and at most one $\sigma'$ leg, and no vertex
carries both. In particular,
\begin{equation}
V_{\lambda'\sigma'\phi}=0,\quad
V_{\lambda'\sigma' h}=0,\quad
V_{\sigma'\sigma'\cdots}=0,\quad
V_{\lambda'\lambda'\cdots}=0,\quad
V_{h\sigma'\sigma'}=0,\quad
V_{\lambda'\bar c\,c}=V_{\sigma'\bar c\,c}=0.
\label{eq:absent-vertices}
\end{equation}

\section{Verification of the matching system and gauge-independence}
\label{app:matching}

In this appendix we derive the coefficients $c_1, c_2, c_3, c_4$
of the LM field renormalization ansatz \eqref{eq:lambda-renorm}
and verify that the on-shell residual equals $203/320$
independently of the gauge parameters $\xi$ and $\zeta$.

\subsection{Matching coefficients}

Comparing $\mathcal{L}_{\rm div}$ \eqref{eq:Ldiv} with
$\mathcal{L}_{\rm CT}$ \eqref{eq:LCT} structure by structure
gives four equations with unique solution:
\begin{align}
c_1 &= -\mathcal{C}_1,
\label{eq:app-c1}
\\
c_2 &= \frac{\mathcal{C}_1}{2}+\mathcal{C}_2,
\label{eq:app-c2}
\\
c_3 &= \frac{\mathcal{C}_1}{2}-\mathcal{C}_4,
\label{eq:app-c3}
\\
c_4 &= -\frac{\mathcal{C}_1}{4}
-\frac{\mathcal{C}_2}{2}
+\mathcal{C}_3
+\frac{\mathcal{C}_4}{2},
\label{eq:app-c4}
\end{align}
in agreement with Eqs.~\eqref{eq:c1}--\eqref{eq:c4} of the main
text. Substituting $c_3$ and $c_4$ into the coefficient of
$(\partial\bar\phi)^4$ in Eq.~\eqref{eq:Ldiv} yields
\begin{equation}
-\frac{c_3}{4}+\frac{c_4}{2}
=
-\frac{\mathcal{C}_1}{4}
-\frac{\mathcal{C}_2}{4}
+\frac{\mathcal{C}_3}{2}
+\frac{\mathcal{C}_4}{2},
\label{eq:app-phi4-coeff}
\end{equation}
so the residual in the $\kappa^4(\partial\bar\phi)^4$ coefficient
between $\mathcal{L}_{\rm CT}$ and $\mathcal{L}_{\rm div}$ is
\begin{equation}
\kappa^4\Bigl[\mathcal{C}_5 - \Bigl(-\frac{\mathcal{C}_1}{4}
-\frac{\mathcal{C}_2}{4}
+\frac{\mathcal{C}_3}{2}
+\frac{\mathcal{C}_4}{2}\Bigr)\Bigr]
=
\kappa^4\Bigl(
\frac{\mathcal{C}_1}{4}+\frac{\mathcal{C}_2}{4}
-\frac{\mathcal{C}_3}{2}-\frac{\mathcal{C}_4}{2}+\mathcal{C}_5
\Bigr)
= \kappa^4\Omega,
\label{eq:app-residual}
\end{equation}
where $\Omega$ is defined in Eq.~\eqref{eq:Delta-def}.

\subsection{Gauge-independence: \texorpdfstring{$\Omega=203/320$}{Omega = 203/320}}

The on-shell combination $\Omega \equiv \mathcal{C}_1/4
+\mathcal{C}_2/4 -\mathcal{C}_3/2-\mathcal{C}_4/2+\mathcal{C}_5$
also equals the on-shell value of
$\mathcal{L}_{\rm CT}$ (obtained by substituting
$\bar R_{\mu\nu}\to-\frac{\kappa^2}{2}\partial_\mu\bar\phi\,
\partial_\nu\bar\phi$, $\bar R\to-\frac{\kappa^2}{2}(\partial\bar\phi)^2$,
$\Dbar^2\bar\phi\to 0$). Writing $x=\xi-1$ and
$y=\xi-\zeta$, the five contributions expand as
\begin{align}
\tfrac{\mathcal{C}_1}{4} &= \tfrac{43}{480}
  +\tfrac{x^2+x}{12},
&
\tfrac{\mathcal{C}_2}{4} &= \tfrac{1}{320}+\tfrac{x^2}{24},
\nonumber\\
-\tfrac{\mathcal{C}_3}{2} &= \tfrac{1}{24}
  +\tfrac{x^2+6x}{24}-\tfrac{y^2}{8},
&
-\tfrac{\mathcal{C}_4}{2} &= -\tfrac{x(2x+7)}{12},
\nonumber\\
\mathcal{C}_5 &= \tfrac{1}{2}+\tfrac{x}{4}+\tfrac{y^2}{8}.
&&
\label{eq:app-expansion}
\end{align}
The coefficients of $y^2$, $x^2$, and $x$ all vanish identically:
\begin{equation}
[y^2]:\;{-\tfrac{1}{8}+\tfrac{1}{8}=0},\quad
[x^2]:\;{\tfrac{2+1+1-4}{24}=0},\quad
[x]:\;{\tfrac{1+3+3-7}{12}=0}.
\end{equation}
The constant term gives
\begin{equation}
\Omega
=
\frac{43}{480}+\frac{1}{320}+\frac{1}{24}+\frac{1}{2}
=
\frac{86+3+40+480}{960}
=
\frac{609}{960}
=
\frac{203}{320},
\label{eq:app-constant}
\end{equation}
completing the proof of Eq.~\eqref{eq:onshell-CT}.

\section{The \texorpdfstring{$\phi^3$}{phi3} model with a Lagrange multiplier field}
\label{app:phi3}

This appendix illustrates, in the simpler setting of a self-coupled
scalar field theory, the same mechanism by which LM renormalization
fails to absorb all one-loop divergences. The model is described by
the Lagrangian
\begin{equation}
\mathcal{L}
=
\tfrac{1}{2}\partial^\mu\phi\,\partial_\mu\phi
+
\frac{g}{3!}\,\phi^3,
\label{eq:phi3-lag}
\end{equation}
where the coupling constant $g$ is dimensionless in $D=6$ spacetime
dimensions. This model involves no gauge structure and thus provides a
clean illustration of the basic features without additional
complications~\cite{Muta:1987vi}.

We employ the background-field method~\cite{Abbott:1981ke}, splitting
\begin{equation}
\phi = \bar\phi + \varphi,
\label{eq:phi3-split}
\end{equation}
where $\bar\phi$ is the background field and $\varphi$ is the quantum
fluctuation. Expanding Eq.~\eqref{eq:phi3-lag} gives
\begin{align}
\mathcal{L}
&=
\tfrac{1}{2}\partial_\mu\bar\phi\,\partial^\mu\bar\phi
+\partial_\mu\bar\phi\,\partial^\mu\varphi
+\tfrac{1}{2}\partial_\mu\varphi\,\partial^\mu\varphi
\nonumber\\
&\quad
+
\frac{g}{3!}
\bigl(\bar\phi^3+3\bar\phi^2\varphi+3\bar\phi\varphi^2+\varphi^3\bigr).
\label{eq:phi3-expanded}
\end{align}
A Lagrange multiplier field $\lambda$ is introduced by coupling it to
the functional derivative $\delta\mathcal{L}/\delta\varphi$,
\begin{equation}
\lambda\,\frac{\delta\mathcal{L}}{\delta\varphi}
=
\lambda\!\left(-\partial^2\bar\phi+\frac{g}{2}\bar\phi^2\right)
+\partial_\mu\lambda\,\partial^\mu\varphi
+\frac{g}{2}\lambda\,(2\bar\phi\varphi+\varphi^2).
\label{eq:phi3-lm}
\end{equation}
For vanishing quantum fields ($\varphi=0$), the Euler--Lagrange equation
for $\lambda$ implies the classical equation of motion
\begin{equation}
\partial^2\bar\phi - \frac{g}{2}\bar\phi^2 = 0.
\label{eq:phi3-eom}
\end{equation}

Using Eqs.~\eqref{eq:phi3-expanded} and~\eqref{eq:phi3-lm}, and
omitting terms linear in the quantum field $\varphi$ as required by the
background-field method, the total Lagrangian becomes
\begin{align}
\mathcal{L}_{\rm tot}
&=
\tfrac{1}{2}\partial_\mu\bar\phi\,\partial^\mu\bar\phi
+\lambda\!\left(-\partial^2\bar\phi+\frac{g}{2}\bar\phi^2\right)
+\tfrac{1}{2}\partial_\mu\varphi\,\partial^\mu\varphi
+\partial_\mu\lambda\,\partial^\mu\varphi
\nonumber\\
&\quad
+\frac{g}{3!}\bigl(\bar\phi^3+3\bar\phi\varphi^2+\varphi^3\bigr)
+\frac{g}{2}\lambda\,(2\bar\phi\varphi+\varphi^2).
\label{eq:phi3-ltot}
\end{align}

The LM field $\lambda$ is split into a quantum component $\lambda'$ and
a background component $\bar\lambda$,
\begin{equation}
\lambda = \lambda' + \bar\lambda,
\qquad
\bar\lambda = c\,g^2\,\bar\phi,
\label{eq:phi3-lm-bgsplit}
\end{equation}
where $\lambda'$ is a quantum field and $\bar\lambda$ is a background
field, with $c$ to be determined. This split is only a convenient
toy-model representation of the same background-dependent LM
renormalization used in the main text; there, the corresponding effect
is written as a shift of the quantum LM fluctuation, with vanishing
background LM fields.
Substituting the split~\eqref{eq:phi3-lm-bgsplit} into
\eqref{eq:phi3-ltot}, we now project onto the purely background sector
in order to identify the induced background counterterm. Thus we set
\(\varphi=0\) and \(\lambda'=0\). This projection does not discard the
quantum fields from the theory; the additional vertices containing
quantum fields are analyzed separately below. The Lagrangian then
reduces to
\begin{equation}
\mathcal{L}_{\rm tot}\big|_{\varphi=\lambda'=0}
=
\tfrac{1}{2}\partial_\mu\bar\phi\,\partial^\mu\bar\phi
+\frac{g}{3!}\,\bar\phi^3
+\bar\lambda\!\left(-\partial^2\bar\phi+\frac{g}{2}\bar\phi^2\right).
\label{eq:phi3-ltot-bg}
\end{equation}
This leads to the following equation of motion for \(\bar\phi\) in the presence of \(\bar\lambda\) given in Eq. \eqref{eq:phi3-lm-bgsplit}.
\begin{equation}
\partial^2\bar\phi - \frac{g}{2}\bar\phi^2
=\frac{3 c}{2}\,g^3\,\bar\phi^2 - 2c\,g^2\,\partial^2\bar\phi .
\label{eq:phi3-eom-modified}
\end{equation}
The two terms on the right-hand side have the form of the variations
of one-loop cubic-vertex and self-energy corrections, respectively.
The classical equation of motion for \(\bar\phi\) is therefore effectively
unchanged at tree level, confirming the self-consistency of the
background-field split~\eqref{eq:phi3-lm-bgsplit}.
After substituting the split~\eqref{eq:phi3-lm-bgsplit} into
Eq.~\eqref{eq:phi3-ltot}, one observes that the two-point sector
in the $(\varphi,\lambda')$ subsector mixes $\varphi$ and
$\lambda'$ but contains no quadratic term in $\lambda'$ alone. The
matrix propagator in the $(\varphi,\lambda')$ sector is therefore
\begin{equation}
\Delta
=
\begin{pmatrix}
0 & 1/\partial^2 \\[4pt]
1/\partial^2 & -1/\partial^2
\end{pmatrix}.
\label{eq:phi3-prop}
\end{equation}
There is no $\varphi$-$\varphi$ propagator; the $\varphi$-$\lambda'$ propagator
is mixed, and the $\lambda'$-$\lambda'$ propagator carries opposite sign.
As a consequence of this LM propagator structure, and apart from the
counterterm insertions discussed below, no ordinary radiative
corrections beyond one loop are generated.

Using the propagators~\eqref{eq:phi3-prop} and the vertices of
Eq.~\eqref{eq:phi3-ltot}, the one-loop divergences are (with
$\epsilon=(6-D)/2$)
\begin{equation}
T_{\rm div}
=
\frac{g^2}{12(4\pi)^3\epsilon}\,
\tfrac{1}{2}\partial_\mu\bar\phi\,\partial^\mu\bar\phi
\;+\;
\frac{g^3}{12(4\pi)^3\epsilon}\,\bar\phi^3.
\label{eq:phi3-tdiv}
\end{equation}
The scalar self-energy divergence (first term) can be cancelled by
choosing the background LM field~\eqref{eq:phi3-lm-bgsplit} with
\begin{equation}
\bar\lambda = \mathcal{B}\,\bar\phi,
\qquad
\mathcal{B} \equiv c\,g^2 = -\frac{g^2}{24(4\pi)^3\epsilon},
\label{eq:phi3-lm-renorm}
\end{equation}
as follows from the first term on the right-hand side of
Eq.~\eqref{eq:phi3-tdiv}.
However, the second term of Eq.~\eqref{eq:phi3-lm}
then induces the additional divergent contribution
\begin{equation}
V^{\rm div}_{\lambda{\rm -shift}}
= \frac{\mathcal{B}g}{2}\,\bar\phi^3
= -\frac{g^3}{48(4\pi)^3\epsilon}\,\bar\phi^3,
\label{eq:phi3-induced}
\end{equation}
which is opposite in sign and differs by a factor of four from the
$\bar\phi^3$ term in Eq.~\eqref{eq:phi3-tdiv}.
This background choice is therefore not sufficient to cancel all
divergences: a residual $\bar\phi^3$ divergence, not removable by any
LM field redefinition, remains
(The remaining difference between Eqs.~\eqref{eq:phi3-tdiv}
and~\eqref{eq:phi3-induced} can be absorbed by an appropriate
renormalization of the coupling constant \(g\)).

\begin{figure}[ht]
\centering
\begin{tikzpicture}[baseline=(current bounding box.center),
                    line cap=round, line join=round]
  \coordinate (V1) at (-1.1,0);
  \coordinate (V2) at ( 1.1,0);
  \draw[bscalar] (-2.4,0) -- (V1);
  \draw[bscalar] (V2) -- (2.4,0);
  \draw[qscalar] (180:1.1) arc[start angle=180, end angle=90, radius=1.1];
  \draw[qslm]    ( 90:1.1) arc[start angle=90,  end angle=0,  radius=1.1];
  \draw[qscalar] (180:1.1) arc[start angle=180, end angle=270, radius=1.1];
  \draw[qslm]    (270:1.1) arc[start angle=270, end angle=360, radius=1.1];
  \filldraw (V1) circle (1.8pt);
  \filldraw (V2) circle (1.8pt);
  \node[above=1pt] at (-0.78, 0.78) {\small$\varphi$};
  \node[above=1pt] at ( 0.78, 0.78) {\small$\lambda'$};
  \node[below=1pt] at (-0.78,-0.78) {\small$\varphi$};
  \node[below=1pt] at ( 0.78,-0.78) {\small$\lambda'$};
  \node[above=2pt] at (-1.9,0) {\small$\bar\phi$};
  \node[above=2pt] at ( 1.9,0) {\small$\bar\phi$};
  \node at (0,-1.8) {$(a)$};
\end{tikzpicture}
\hspace{2.5cm}
\begin{tikzpicture}[baseline=(current bounding box.center),
                    line cap=round, line join=round]
  \coordinate (A)   at (-1.3,  0.0);
  \coordinate (B)   at ( 0.7,  0.9);
  \coordinate (C)   at ( 0.7, -0.9);
  \coordinate (MAB) at ($(A)!0.5!(B)$);
  \coordinate (MAC) at ($(A)!0.5!(C)$);
  \coordinate (MBC) at ($(B)!0.5!(C)$);
  \draw[bscalar] (-2.5, 0.0) -- (A);
  \draw[bscalar] (B) -- (2.0, 0.9);
  \draw[bscalar] (C) -- (2.0,-0.9);
  \draw[qscalar] (A)   -- (MAB);
  \draw[qslm]    (MAB) -- (B);
  \draw[qscalar] (A)   -- (MAC);
  \draw[qslm]    (MAC) -- (C);
  \draw[qscalar] (B)   -- (MBC);
  \draw[qslm]    (MBC) -- (C);
  \filldraw      (A) circle (1.8pt);
  \filldraw  (B) circle (1.8pt);
  \filldraw  (C) circle (1.8pt);
  \node[above left=0pt]  at ($(A)!0.25!(B)$) {\small$\varphi$};
  \node[above right=5pt] at ($(A)!0.55!(B)$) {\small$\lambda'$};
  \node[below left=0pt]  at ($(A)!0.25!(C)$) {\small$\varphi$};
  \node[below right=5pt] at ($(A)!0.55!(C)$) {\small$\lambda'$};
  \node[right=2pt] at ($(B)!0.25!(C)$) {\small$\varphi$};
  \node[right=2pt] at ($(B)!0.75!(C)$) {\small$\lambda'$};
  \node[above=2pt] at (-1.9, 0.0) {\small$\bar\phi$};
  \node[above=2pt] at ( 1.5, 0.9) {\small$\bar\phi$};
  \node[below=2pt] at ( 1.5,-0.9) {\small$\bar\phi$};
  \node at (-0.3,-1.6) {$(b)$};
\end{tikzpicture}
\caption{Representative one-loop graphs that could arise from the new
vertex induced by the background LM choice~\eqref{eq:phi3-lm-renorm}. Line conventions are as in Appendix~\ref{app:diagrams}: solid and
dash-dot segments denote, respectively, the \(\varphi\) and \(\lambda'\)
ends of mixed propagators.
In both cases the new vertex forces the internal lines to be mixed
$\varphi$-$\lambda'$ propagators (Eq.~\eqref{eq:phi3-prop}), which in
turn requires one of the remaining vertices to carry two $\lambda'$
legs; no such vertex exists in the LM theory.  Both diagrams therefore
vanish identically.}
\label{fig:phi3-graphs}
\end{figure}
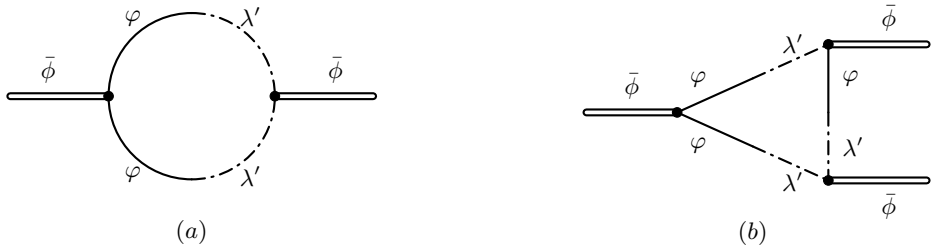

Apart from terms linear in $\varphi$, the
background choice~\eqref{eq:phi3-lm-renorm} also induces in
Eq.~\eqref{eq:phi3-lm} the new vertex
$\frac{\mathcal{B}\,g}{2}\,\bar\phi\varphi^2$ (which also appears,
with coefficient $g/2$, in Eq.~\eqref{eq:phi3-expanded}). One may wonder
whether this vertex generates additional divergent diagrams at one
loop. To see that it does not, consider the graphs in
Fig.~\ref{fig:phi3-graphs}. Since there is no $\varphi$-$\varphi$
propagator, any loop involving the new vertex must be closed using
mixed $\varphi$-$\lambda'$ propagators. This forces at least one remaining
vertex in the diagram to be quadratic in $\lambda'$. Since no such
vertex exists in the LM theory, the contributions from all graphs of
this type vanish identically.

This observation has a further consequence. The induced vertex
$\frac{\mathcal{B}\,g}{2}\,\bar\phi\varphi^2$ is proportional to the
vertex $\frac{g}{2}\bar\phi\varphi^2$ that already appears in the
original, unrenormalized LM Lagrangian~\eqref{eq:phi3-ltot}.
Since the unrenormalized LM theory admits no radiative corrections
beyond one loop, contributions from this type of vertex already
cancel at two loops and higher in the original theory.
The induced vertex, being the same operator rescaled by the one-loop
coefficient $\mathcal{B}$, therefore cannot generate new multi-loop
contributions either. The only effect of the background LM field
choice~\eqref{eq:phi3-lm-renorm} occurs at one loop, where it absorbs
the scalar self-energy divergence~\eqref{eq:phi3-tdiv}.

This is the $\phi^3$ analogue of the result established in
Sec.~\ref{sec:obstruction}: in both cases, LM renormalization absorbs
all divergences except a single residual whose on-shell value is
nonzero, and which therefore no equation-of-motion counterterm can
cancel. In the $\phi^3$ model the residual is the
$\bar\phi^3$ operator; in the gravity-coupled scalar theory it is the
gauge-invariant combination
$\kappa^4(\partial_\mu\bar\phi\,\partial^\mu\bar\phi)^2$.

The extension of this argument to the gravitational and mixed sectors
of the full theory, including the Faddeev--Popov ghost sector, is
given in Sec.~\ref{sec:CT}.

\section{Two-loop illustration: both LM fields are required in the presence of matter}
\label{app:twoloop}

Ref.~\cite{Chishtie:2025} argues that a single tensor LM field
$\lambda'_{\mu\nu}$, coupled to matter as in Eq.~\eqref{eq:LLM} but
without the scalar multiplier $\sigma'$, already eliminates all
radiative corrections beyond one loop, even in the presence of matter,
and uses this claim to argue for a consistent unification of the LM
formulation of gravity with the Standard Model. This appendix shows,
by explicit two-loop examples, that this is not the case: with
$\lambda'_{\mu\nu}$ alone, genuine two-loop divergences survive in
graphs that mix matter and gravitational lines. The scalar multiplier
$\sigma'$ is not an optional refinement; it is required to enforce
one-loop exactness once matter is present, while the irreducible one-loop
residual $\kappa^4(203/320)(\partial\bar\phi)^4$ remains, as discussed
in Sec.~\ref{sec:obstruction}. Line conventions are as in
Appendix~\ref{app:diagrams}.

\subsection{Pure gravity: \texorpdfstring{$\lambda'_{\mu\nu}$}{lambda'} alone already suffices}

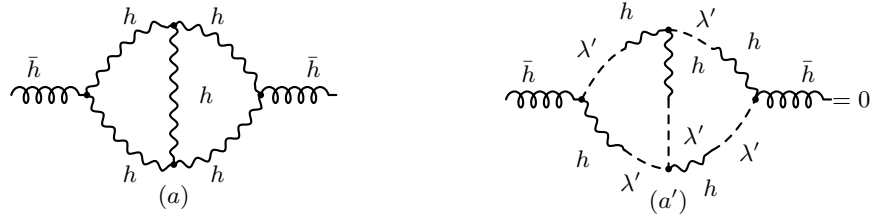
\begin{figure}[ht]
\centering
\begin{tikzpicture}[baseline=(current bounding box.center),
                    line cap=round,line join=round]
  \coordinate (L) at (-1.15,0);
  \coordinate (R) at ( 1.15,0);
  \coordinate (T) at (0,0.92);
  \coordinate (B) at (0,-0.92);
  \coordinate (Lext) at (-2.15,0);
  \coordinate (Rext) at ( 2.15,0);
  \draw[bgrav] (Lext)--(L);
  \draw[bgrav] (R)--(Rext);
  \draw[qgrav] (L) .. controls (-0.72,0.70) and (-0.36,0.92) .. (T);
  \draw[qgrav] (T) .. controls ( 0.36,0.92) and ( 0.72,0.70) .. (R);
  \draw[qgrav] (L) .. controls (-0.72,-0.70) and (-0.36,-0.92) .. (B);
  \draw[qgrav] (B) .. controls ( 0.36,-0.92) and ( 0.72,-0.70) .. (R);
  \draw[qgrav] (T)--(B);
  \vertexdot{L}\vertexdot{R}\vertexdot{T}\vertexdot{B}
  \node[above] at (-1.85,0.12) {\small$\bar h$};
  \node[above] at ( 1.85,0.12) {\small$\bar h$};
  \node[above=4pt] at (-0.58,0.68) {\small$h$};
  \node[above=4pt] at ( 0.58,0.68) {\small$h$};
  \node[below=4pt] at (-0.58,-0.68) {\small$h$};
  \node[below=4pt] at ( 0.58,-0.68) {\small$h$};
  \node[right=4pt] at (0.08,0) {\small$h$};
  \node at (0,-1.35) {$(a)$};
\end{tikzpicture}
\hspace{2.0cm}
\begin{tikzpicture}[baseline=(current bounding box.center),
                    line cap=round,line join=round]
  \coordinate (L) at (-1.15,0);
  \coordinate (R) at ( 1.15,0);
  \coordinate (T) at (0,0.92);
  \coordinate (B) at (0,-0.92);
  \coordinate (ULT) at (-0.58,0.67);
  \coordinate (URT) at ( 0.58,0.67);
  \coordinate (LLB) at (-0.58,-0.67);
  \coordinate (LRB) at ( 0.58,-0.67);
  \coordinate (Lext) at (-2.15,0);
  \coordinate (Rext) at ( 2.15,0);
  \draw[bgrav] (Lext)--(L);
  \draw[bgrav] (R)--(Rext);
  \draw[qlm]   (L) .. controls (-0.94,0.40) and (-0.76,0.58) .. (ULT);
  \draw[qgrav] (ULT) .. controls (-0.38,0.82) and (-0.18,0.92) .. (T);
  \draw[qlm]   (T) .. controls (0.18,0.92) and (0.38,0.82) .. (URT);
  \draw[qgrav] (URT) .. controls (0.76,0.58) and (0.94,0.40) .. (R);
  \draw[qgrav] (L) .. controls (-0.94,-0.40) and (-0.76,-0.58) .. (LLB);
  \draw[qlm]   (LLB) .. controls (-0.38,-0.82) and (-0.18,-0.92) .. (B);
  \draw[qgrav] (B) .. controls (0.18,-0.92) and (0.38,-0.82) .. (LRB);
  \draw[qlm]   (LRB) .. controls (0.76,-0.58) and (0.94,-0.40) .. (R);
  \mixedWDc{T}{B}
  \vertexdot{L}\vertexdot{R}\vertexdot{T}\vertexdot{B}
  \node[above left]  at (-0.82,0.45) {\small$\lambda'$};
  \node[above left=3pt]  at (-0.24,0.84) {\small$h$};
  \node[above right] at ( 0.24,0.84) {\small$\lambda'$};
  \node[above right=3pt] at ( 0.82,0.45) {\small$h$};
  \node[below left=3pt]  at (-0.82,-0.45) {\small$h$};
  \node[below left]  at (-0.24,-0.84) {\small$\lambda'$};
  \node[below right=3pt] at ( 0.24,-0.84) {\small$h$};
  \node[below right] at ( 0.82,-0.45) {\small$\lambda'$};
  \node[right=3pt] at (0.08, 0.46) {\small$h$};
  \node[right] at (0.08,-0.46) {\small$\lambda'$};
  \node[above] at (-1.85,0.12) {\small$\bar h$};
  \node[above] at ( 1.85,0.12) {\small$\bar h$};
  \node[right] at (2.05,0) {$=0$};
  \node at (0,-1.35) {$(a')$};
\end{tikzpicture}
\caption{Two-loop pure-gravity graphs. (a): built from the ordinary
graviton propagator, which vanishes identically in the LM formulation,
so this graph does not occur. (a$'$): the corresponding graph built
from the propagators that actually exist, using only $\lambda'_{\mu\nu}$;
this requires a vertex quadratic in $\lambda'_{\mu\nu}$, which does not
exist, so the graph vanishes.}
\label{fig:twoloop-b}
\end{figure}

Consider first a two-loop, purely gravitational graph built from
ordinary graviton lines, Fig.~\ref{fig:twoloop-b}(a). This vanishes
trivially, since the ordinary graviton propagator
$\langle h_{\mu\nu}h_{\rho\sigma}\rangle_q$ is identically zero in the
LM formulation (Sec.~\ref{sec:formulation}); no diagram built from it
can occur. The corresponding graph built from the propagators that
actually exist in the theory, using only the mixed $\langle
h\lambda'\rangle_q$ propagator throughout, Fig.~\ref{fig:twoloop-b}(a$'$),
requires a vertex quadratic in $\lambda'_{\mu\nu}$ to close the
diagram; no such vertex exists in the LM theory (Sec.~\ref{sec:CT}),
so this graph vanishes as well. Both graphs are therefore absent, and
$\lambda'_{\mu\nu}$ alone is sufficient to enforce one-loop exactness
in the pure-gravity sector, consistent with
Sec.~\ref{sec:puregravity}.

\subsection{Matter: \texorpdfstring{$\sigma'$}{sigma'} is required}

\begin{figure}[ht]
\centering
\begin{tikzpicture}[baseline=(current bounding box.center),
                    line cap=round,line join=round]
  \coordinate (L) at (-1.15,0);
  \coordinate (R) at ( 1.15,0);
  \coordinate (T) at (0,0.92);
  \coordinate (B) at (0,-0.92);
  \coordinate (Lext) at (-2.15,0);
  \coordinate (Rext) at ( 2.15,0);
  \draw[bscalar] (Lext)--(L);
  \draw[bscalar] (R)--(Rext);
  \draw[qscalar] (L)--(R);
  \draw[qlm]   (L) .. controls (-0.72,0.70) and (-0.36,0.92) .. (T);
  \draw[qgrav] (T) .. controls ( 0.36,0.92) and ( 0.72,0.70) .. (R);
  \draw[qgrav] (L) .. controls (-0.72,-0.70) and (-0.36,-0.92) .. (B);
  \draw[qlm]   (B) .. controls ( 0.36,-0.92) and ( 0.72,-0.70) .. (R);
  \vertexdot{L}\vertexdot{R}
  \node[above] at (-1.85,0.08) {\small$\bar\phi$};
  \node[above] at ( 1.85,0.08) {\small$\bar\phi$};
  \node[above] at (-0.58,0.68) {\small$\lambda'$};
  \node[above] at ( 0.58,0.68) {\small$h$};
  \node[below] at (-0.58,-0.68) {\small$h$};
  \node[below] at ( 0.58,-0.68) {\small$\lambda'$};
  \node[above] at (0,0.05) {\small$\varphi$};
  \node at (0,-1.35) {$(a)$};
\end{tikzpicture}
\hspace{2.0cm}
\begin{tikzpicture}[baseline=(current bounding box.center),
                    line cap=round,line join=round]
  \coordinate (L) at (-1.15,0);
  \coordinate (R) at ( 1.15,0);
  \coordinate (T) at (0,0.92);
  \coordinate (B) at (0,-0.92);
  \coordinate (Lext) at (-2.15,0);
  \coordinate (Rext) at ( 2.15,0);
  \draw[bscalar] (Lext)--(L);
  \draw[bscalar] (R)--(Rext);
  \mixedSscalarc{L}{R}
  \draw[qlm]   (L) .. controls (-0.72,0.70) and (-0.36,0.92) .. (T);
  \draw[qgrav] (T) .. controls ( 0.36,0.92) and ( 0.72,0.70) .. (R);
  \draw[qgrav] (L) .. controls (-0.72,-0.70) and (-0.36,-0.92) .. (B);
  \draw[qlm]   (B) .. controls ( 0.36,-0.92) and ( 0.72,-0.70) .. (R);
  \vertexdot{L}\vertexdot{R}
  \node[above] at (-1.85,0.08) {\small$\bar\phi$};
  \node[above] at ( 1.85,0.08) {\small$\bar\phi$};
  \node[above] at (-0.58,0.68) {\small$\lambda'$};
  \node[above] at ( 0.58,0.68) {\small$h$};
  \node[below] at (-0.58,-0.68) {\small$h$};
  \node[below] at ( 0.58,-0.68) {\small$\lambda'$};
  \node[above] at (-0.58,0.05) {\small$\varphi$};
  \node[above] at ( 0.58,0.05) {\small$\sigma'$};
  \node[right] at (2.05,0) {$=0$};
  \node at (0,-1.35) {$(a')$};
\end{tikzpicture}
\caption{Two-loop mixed matter--gravity graphs. (a): built from
$\lambda'_{\mu\nu}$ alone, with the ordinary matter propagator
$\langle\varphi\varphi\rangle_q$ internally; this graph is topologically
admissible and is not eliminated by the LM structure. (a$'$): the corresponding
graph once $\sigma'$ is introduced, with the mixed
$\langle\varphi\sigma'\rangle_q$ propagator internally; this requires a
vertex containing both $\lambda'_{\mu\nu}$ and $\sigma'$, which does not
exist, so the graph vanishes.}
\label{fig:twoloop-a}
\end{figure}
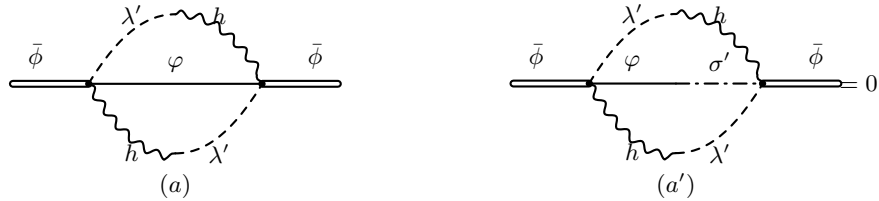

Now consider the analogous two-loop graph mixing matter and
gravitational lines, using $\lambda'_{\mu\nu}$ alone (no $\sigma'$),
Fig.~\ref{fig:twoloop-a}(a). Unlike the pure-gravity case, this graph
does \emph{not} vanish: the internal $\langle\varphi\varphi\rangle_q$
propagator is the ordinary, non-vanishing matter propagator, since no
mechanism in the theory --- absent $\sigma'$ --- forces it to zero, and
the vertices required to build the diagram (coupling $\lambda'_{\mu\nu}$
simultaneously to $h_{\mu\nu}$ and to matter) are present in the theory.
This diagram is therefore topologically admissible and is not
eliminated by the LM propagator structure: the combinatorial argument
of Sec.~\ref{app:twoloop-allorder} (the bound $L\le1-V_0$) simply does
not apply here, since $\lambda'_{\mu\nu}$ alone leaves
$\langle\varphi\varphi\rangle_q$ nonzero. One-loop exactness in the
presence of matter is therefore not guaranteed by $\lambda'_{\mu\nu}$
alone, contrary to the claim of Ref.~\cite{Chishtie:2025} that a
single LM field suffices to eliminate all higher-loop matter
corrections.

Once the scalar multiplier $\sigma'$ is introduced, the ordinary
$\langle\varphi\varphi\rangle_q$ propagator is replaced by the mixed
$\langle\varphi\sigma'\rangle_q$ propagator (Appendix~\ref{app:diagrams}),
giving the graph of Fig.~\ref{fig:twoloop-a}(a$'$). This graph requires
a vertex containing both $\lambda'_{\mu\nu}$ and $\sigma'$ simultaneously;
since $\lambda'_{\mu\nu}$ and $\sigma'$ are coupled, respectively, only
to the gravitational and to the scalar equation of motion
(Sec.~\ref{sec:formulation}), no such mixed vertex exists, and the graph
vanishes identically.

The contrast between Figs.~\ref{fig:twoloop-b} and~\ref{fig:twoloop-a}
is the diagrammatic content of the obstruction discussed in
Sec.~\ref{sec:obstruction}: in the pure-gravity sector,
$\lambda'_{\mu\nu}$ alone already forbids every graph beyond one loop,
but once matter is present, only the combination of $\lambda'_{\mu\nu}$
\emph{and} $\sigma'$ forbids the corresponding graphs, since it is
specifically the absence of any $\lambda'$-$\sigma'$ vertex --- not the
absence of a $\lambda'\lambda'$ vertex, which is already exploited by
$\lambda'_{\mu\nu}$ alone in the pure-gravity case --- that removes the
particular graphs shown.

\subsection{All-order generalization}
\label{app:twoloop-allorder}

The extension to arbitrary loop order follows from a topological
counting argument.  Write the complete LM system in field-space notation
as
\begin{equation}
 S_{\rm tot}[Q,\Lambda]
 =S[Q]+\Lambda_A\frac{\delta S[Q]}{\delta Q^A},
 \qquad
 Q^A=(h_{\mu\nu},\varphi),\qquad
 \Lambda_A=(\lambda'_{\mu\nu},\sigma'),
\label{eq:all-order-LM-action}
\end{equation}
where $S$ denotes the full gauge-fixed action in the physical-field
sector.  The inverse of the quadratic form has no $QQ$ block,
$\langle Q^A Q^B\rangle_q=0$, whereas mixed $Q\Lambda$ and
$\Lambda\Lambda$ propagators may be nonzero.  Moreover, every vertex
arising from the second term in Eq.~\eqref{eq:all-order-LM-action}
contains exactly one LM leg, and an ordinary vertex arising from $S[Q]$
contains none.

Consider a connected graph built from the $(Q,\Lambda)$ fields and
contributing to the effective action, with external legs only in the
physical/background sector.  Let $V_0$ and $V_1$ denote, respectively,
the numbers of ordinary vertices and single-LM vertices, and let $I$ be
the number of internal lines.  Since
the $QQ$ propagator vanishes, every internal line must have at least one
LM endpoint.  All LM legs are internal in the class of graphs under
consideration, and there are precisely $V_1$ such endpoints.  Therefore
\begin{equation}
 I\leq V_1.
\label{eq:all-order-line-bound}
\end{equation}
Using the topological identity for the number of loops of a connected
graph gives
\begin{equation}
 L=I-(V_0+V_1)+1
 \leq 1-V_0
 \leq 1.
\label{eq:all-order-loop-bound}
\end{equation}
Thus no connected contribution beyond one loop can occur when an LM
partner is introduced for every quantum field in the coupled system,
within the $(Q,\Lambda)$ vertex/line counting above. The
Faddeev--Popov and Lee--Yang sectors are not part of that counting;
their treatment as the functional determinants discussed in
Sec.~\ref{sec:formulation} is an additional, separate assumption. Since
both sectors are bilinear in their respective ghost fields, they
supply only the corresponding one-loop determinants and do not
introduce further loop corrections, so the localization argument is
unaffected once this is granted.

This counting also shows why $\lambda'_{\mu\nu}$ alone is insufficient
in the presence of matter.  Without $\sigma'$, the scalar propagator
$\langle\varphi\varphi\rangle_q$ is nonzero, so a scalar internal line
need not consume an LM endpoint and the bound~\eqref{eq:all-order-line-bound}
no longer follows.  Mixed gravity--matter graphs with $L>1$, such as
Fig.~\ref{fig:twoloop-a}(a), are then allowed.  The requirement that
both LM fields be present once matter is coupled is therefore not a
technical inconvenience; it is the mechanism that renders the complete
coupled theory one-loop exact, while leaving the irreducible
residual~\eqref{eq:Ladd}.


\end{document}